\documentclass [12pt,fleqn]{article}

\usepackage{verbatim}
\usepackage{graphicx}
\usepackage{caption}
\usepackage{amssymb}
\usepackage{amsmath}
\usepackage{lscape}
\usepackage{xcolor}
\usepackage{float}

\newtheorem {lemma} {Lemma}
\newtheorem {theorem} {Theorem}
\newtheorem {corollary} {Corollary}

\newenvironment{apabib}{
  \noindent
{\large\bf References} \vspace*{-6mm} 
\begin{list}{}{
  \topsep          0mm
  \leftmargin     10mm
  \parsep         2mm 
  \listparindent -10mm}
  \item[]\strut\par}{
\end{list}}

\newcounter{excount}
\newenvironment{example}
{\refstepcounter{excount}\bigskip\noindent\textbf{Example~\arabic{excount}:}}{$\;\rule{1.5mm}{3mm}$\medskip}

\newcounter{asscount}
\newenvironment{assumption}
{\refstepcounter{asscount}\bigskip\noindent\textsc{Assumption~\arabic{asscount}}}
{\medskip}

\newcounter{defcount}
\newenvironment{definition}
{\refstepcounter{defcount}\bigskip\noindent\textsc{Definition~\arabic{defcount}}}
{\smallskip}

\newcommand{\prf}{\noindent {\bf Proof}:\ \ } 
\newcommand{\eprf}{$\;\rule{1.5mm}{3mm}$ \smallskip}

\newcommand{\mcalP}{\mathcal{P}}
\renewcommand{\mid}{|\,}

\begin{document}

\title{Interacting Treatments with Endogenous Takeup\thanks{We thank Aniko Biro, Matias Cattaneo, Noemi Kreif, Attila Lindner, Laszlo Matyas, Timea Molnar and three anonymous referees for useful comments. All errors are our responsibility. An earlier version of the paper was circulated under the title ``Treatment Effect Analysis for Pairs with Endogenous Treatment Takeup.''}}

\author{M\'at\'e Kormos\footnote{Delft University of Technology, email: m.kormos@tudelft.nl}\and
Robert P.\ Lieli\footnote{Central European University, Vienna, email: lielir@ceu.edu}\and
Martin Huber\footnote{University of Fribourg, email: martin.huber@unifr.ch}}

\maketitle

\thispagestyle{empty}

\begin{abstract}
{\footnotesize

We study causal inference in randomized experiments (or quasi-experiments) following a $2\times 2$ factorial design. There are two treatments, denoted $A$ and $B$, and units are randomly assigned to one of four categories: treatment $A$ alone, treatment $B$ alone, joint treatment, or none. Allowing for endogenous non-compliance with the two binary instruments representing the intended assignment, as well as unrestricted interference across the two treatments, we derive the causal interpretation of various instrumental variable estimands under more general compliance conditions than in the literature. 
In general, if treatment takeup is driven by both instruments for some units, it becomes difficult to separate treatment interaction from treatment effect heterogeneity. We provide auxiliary conditions and various bounding strategies that may help zero in on causally interesting parameters. As an empirical illustration, we apply our results to a program randomly offering two different treatments, namely tutoring and financial incentives, to first year college students, in order to assess the treatments' effects on academic performance.

}



\end{abstract}

\newpage

\section{Introduction}

The experimental approach to establishing the causal effect of a treatment is based on allocating units randomly to treatment and control, thereby precluding any systematic difference between the two groups other than the treatment itself. While unit-level treatment effects may be heterogeneous, comparing the average outcome in the treated and control groups gives a consistent estimate of the average treatment effect. This deceptively simple description of the experimental ideal, which originates in the work of Fisher (1925), embodies several further assumptions, formalized later by Rubin (1974, 1978) and others. A lot of subsequent work on causal inference has sought to extend the analysis of experimental data to more complicated situations with some of the following features.

First, in real world experiments perfect compliance with the intended treatment assignment is not always possible or ethical to enforce. If non-compliance is endogenous (i.e., it depends on unobserved confounders),
then the average difference  between the treated and non-treated values does not represent the treatment effect alone but selection effects as well. Second, randomization itself does not ensure that the treatment status of an individual does not interfere with the potential outcomes of another, violating what is called the stable unit treatment value assumption (SUTVA) in the Rubin causal model. Third, there are experimental setups in which units are potentially subject to multiple, but not mutually exclusive, treatments that may interact with each other.

In this paper we derive the causal interpretation of various instrumental variable (IV) estimands in an experimental or quasi-experimental setup that extends the basic model in all three directions mentioned above. More concretely:

\begin{enumerate}
\item [(i)] 
Population units are targeted by two binary (0/1) treatments, denoted as $D_A$ and $D_B$. 

\item [(ii)] 
The individual treatment effects and the extent to which the two treatments interact may vary from unit to unit in an arbitrary way. 

\item [(iii)] There are two binary (0/1) instruments, denoted as $Z_A$ and $Z_B$, representing randomized assignment to the corresponding treatment or some exogenous incentive to take that treatment. Nevertheless, compliance is imperfect (endogenous), including the possibility of ``instrument spillovers,'' where $Z_A$ affects the takeup of $D_B$ or $Z_B$ affects the takeup of $D_A$. We also refer to this situation as ``coordinated compliance'' (see Example~\ref{ex: couples} below).



\end{enumerate}

We subsequently present two examples which illustrate these ideas.

\begin{example}\label{ex: grades}
Angrist, Lang and Oreopoulos (2009) assess a randomized program providing two treatments to first-year college students: academic services in the form of tutoring ($D_A$) and financial incentives ($D_B$), both aimed at improving performance on end-of-semester exams. Students who entered college 
in September 2005 and had a high school grade point average lower than the upper quartile were randomly offered either one, both, or no treatment. Therefore, the
instruments $Z_A$ and $Z_B$ are binary indicators for being offered tutoring and/or financial incentives. 
While the offers are randomly assigned, the actual takeup of tutoring or financial incentives is likely to be endogenous as it might be driven by personality traits also affecting academic performance. In addition, the treatments may interact such that, for instance, the effectiveness of financial incentives might increase when also receiving tutoring. Section \ref{sec: application} provides an illustration of our results based on this example.
\end{example}

\begin{example}\label{ex: couples}
The population of interest comprises of married couples where one member suffers from depression and the other does not. There are two binary treatments: an antidepressant medication for the depressed spouse ($D_A$), and an educational program about depression for the healthy spouse ($D_B$). The dependent variable may measure the severity of the depression symptoms. Even if the intended treatment assignments ($Z_A$ and $Z_B$) are random, the actual compliance decision may well be endogenous and coordinated across the spouses (for example, they might agree that they will take the treatments if and only if both of them are assigned). Moreover, the two treatments may interact; the medication might be more effective if accompanied by behavioral adjustments on the partner's part.
\end{example}


Motivated by Example~\ref{ex: couples}, it will be convenient in the rest of the paper to represent population units as pairs $(A,B)$, where member $A$ is targeted by $D_A/Z_A$ and member $B$ is targeted by $D_B/Z_B$. The outcome of interest may be be associated with member $A$, member $B$ or the pair itself. Thus, we can equate treatment interaction with interference across pair members (a violation of SUTVA). The representation of units in terms of pairs comes without loss of generality, as a single individual targeted by two treatments (such as in Example~\ref{ex: grades}) can always be thought of as a ``pair'' with identical members.

The estimands we consider in this framework derive from two simple IV regressions and a saturated IV regression. Specifically, we study the causal interpretation of the standard Wald estimand associated with treatment/instrument $A$ conditional on $Z_B=0$ and $Z_B=1$, respectively, as well as the IV (2SLS) regression of the outcome on $D_A$, $D_B$ and $D_AD_B$, instrumented by $Z_A$, $Z_B$ and $Z_AZ_B$. There is now a substantial econometrics literature on causal inference in similar settings. 
Our results contribute to this growing body of work in the following ways.\footnote{In Section \ref{sec: lit} below we provide a brief literature review and position our paper more carefully relative to the most relevant subset of papers.}

First, we employ weaker restrictions on instrument spillovers than other studies in the literature. Specifically, we allow for a compliance type for whom the presence of, say, $Z_B$ represents a strong incentive against taking treatment $A$; we call this type the \emph{cross-defiers}. At the same time, our framework also accommodates \emph{joint compliers}---a type that reacts positively to the presence of the partner instrument and ultimately takes the given treatment if (and only if) both instruments are present. While identification results with joint compliers are available in other studies (e.g., Vazquez-Bare 2022), cross-defiers are typically ruled out, despite our application in Section \ref{sec: application} showing that this is also an empirically relevant type.    

Second, our general identification results make explicit the difficulties that instrument spillovers cause in identifying ``standalone'' average treatment effects and, even more starkly, in separating treatment interaction from treatment effect heterogeneity. In particular, we provide the causal interpretation of the interaction term in the saturated IV regression, and show that a generally interesting local average interaction effect is confounded by terms that depend on how different the average effects of the two treatments are across various subgroups. These confounding terms however vanish if there are no interactive types or there is no treatment effect heterogeneity. 

Third, given the general lack of interpretability of the interaction term in the IV regression, we also provide partial identification results---that is, bounds---for a parameter measuring the average interaction effect of the two treatments in a specific subgroup. We bound the interaction effect directly, based on its definition, as well as indirectly by bounding the confounding heterogeneity terms that show up in the interpretation of the interaction term. We further consider a formal Manski-type bounding strategy, which only uses the data and weak auxiliary assumptions, and a less formal strategy that relies on heuristic restrictions on treatment effect heterogeneity. The application illustrates all bounding approaches. 

Fourth, while the results discussed in the main text impose one-sided noncompliance on the individual instruments (similarly to many results in the literature), we explore the consequences of relaxing this powerful but restrictive assumption in an appendix to the paper.\footnote{One-sided noncompliance means that the treatment cannot be accessed without receiving the instrument.} Specifically, we extend the analysis of the two Wald estimands to the case in which only one of the two instruments satisfies one-sided noncompliance, and even provide causal interpretations that do not require \emph{any} monotonicity conditions. 

The rest of the paper is organized as follows. In Section~\ref{sec: lit} we position our paper in the literature. Section \ref{sec: RCM for pairs} presents a formal potential outcome framework for pairs with endogenous (non-)compliance. We state and discuss our identification results in Section \ref{sec: identification}. Section \ref{sec: application} applies our theory in the context of Example~\ref{ex: grades}. 
Section \ref{sec: concl} concludes. There is a substantial amount of supplementary material relegated to appendices. Appendix~\ref{app: cond mom id} provides supporting identification results. Appendix~\ref{app: relax 1-sided nc} explores the relaxation of one-sided non-compliance. Appendix~\ref{app: proofs main} contains the proofs of all results in the main text, and, finally, Appendix~\ref{app: emp results} supplements the empirical application.

\section{Related literature}\label{sec: lit}

The seminal paper by Imbens and Angrist (1994) investigates endogenous non-compliance with a binary instrument, representing intended assignment to a binary treatment, when individual treatment effects are heterogeneous. 
The study establishes the now well-known result that a simple IV regression identifies the local average treatment effect (LATE) among compliers, the subpopulation whose treatment status complies with the instrument, under conditions 
that include the weak monotonicity of the treatment in the instrument. 

Several studies extend this framework 
to multiple treatments or instruments, which typically entails refinements of the monotonicity assumption. For example, Mogstad, Torgovitsky, and Walters (2020) impose partial monotonicity of the treatment in one instrument conditional on the other instrument(s), while Goff (2022) invokes vector monotonicity, which implies that each instrument affects treatment uptake in a direction that is common across subjects.
Closer to our framework, Behaghel, Cr\'{e}pon and Gurgand (2013) consider two mutually exclusive binary treatments along with two binary instruments. 
Ruling out instrument spillovers, they identify the LATEs among the two groups of compliers. Kirkeboen, Leuven, and Mogstad (2016) exploit information on next-best treatment alternatives to ease the assumption of no instrument spillovers. Heckman and Pinto (2018) allow for instrument spillovers but impose an unordered monotonicity assumption. The latter requires that if some subjects move into (out of) a treatment when the instrument values are switched, then no subjects can at the same time move out of (into) that respective treatment. Lee and Salani\'{e} (2018) discuss LATE identification without an unordered monotonicity assumption in the presence of sufficiently many continuous (rather than binary) instruments when treatment choice is governed by threshold-crossing models. 


Another strand of related literature is concerned with relaxations of SUTVA, allowing for specific forms of interference among treatments, in various (quasi-) experimental settings; see for instance Sobel (2006),  Hong and Raudenbush (2006), Hudgens and Halloran (2008), Ferracci, Jolivet and van den Berg (2014) or Huber and Steinmayr (2021). Particularly relevant in our context are the studies by Kang and Imbens (2016), Imai, Jiang and Malani (2021), and Vazquez-Bare (2022), who combine relaxations of SUTVA with treatment non-compliance. In addition, Blackwell (2017) studies the interaction between two randomized treatments with non-compliance with an application in political science.
We now discuss the relationship between this last set of papers and ours in more detail.


Kang and Imbens (2016) and Imai, Jiang and Malani (2021) consider a partial interference framework, where interference between a given unit's outcome and a peer's treatment occurs within well-defined clusters, such as geographic regions. (In our setting the clusters are the pairs, i.e., there is only one peer.) In addition to conventional IV assumptions, 
Kang and Imbens (2016) impose a treatment exclusion restriction. This assumption posits that each unit's treatment uptake depends solely on their own instrument, but not on the peers' instruments. Ruling out coordinated compliance, 
this restriction permits identification of the direct LATE, i.e., the average effect of the own treatment among compliers. Under an additional one-sided non-compliance assumption, 
one can also identify the average interference (spillover) effect of the peers' treatments, in the absence of the own treatment, 
within the whole population.
Imai, Jiang, and Malani (2021) also study the identification of the direct LATE and present a condition that holds under the treatment exclusion restriction, but is even satisfied under the weaker condition that
a unit's treatment status does not depend on the instrument values of those peers who are non-compliers with respect to their own instruments. Our paper allows for even more general forms of instrument spillovers and makes explicit the resulting difficulties in identifying easily interpretable causal effects. 

The framework of our paper is most closely related to Blackwell (2017; henceforth, BW) and Vazquez-Bare (2022; henceforth, VB), but it extends each in certain directions.\footnote{The basic features of our framework were originally developed in the MA thesis of Kormos (2018).} BW does not explicitly consider (partial) interference, but a scenario where a unit's outcome may be affected by two interacting binary treatments, \( D_A \) and \( D_B \), each with a distinct instrument, \( Z_A \) and \( Z_B \), respectively. As discussed above, this setup can be viewed as a special case of our framework 
where pair members \( A \) and \( B \) are each targeted by a corresponding treatment and instrument, with one unit's treatment potentially having an interference effect on the other unit's outcome. 
Just as Kang and Imbens (2016), BW imposes a treatment exclusion restriction to identify the direct LATE of a given treatment (say, \( D_A \)) conditional on the other treatment (\( D_B = 0,1\)), among complier units 
who follow the respective instruments in both their takeup decisions. Again, our framework and VB's are more general in that instrument spillovers are allowed; on the other hand, BW's results do not impose one-sided noncompliance.

VB explicitly considers a paired design and imposes a weaker first stage condition than the treatment exclusion restriction. Specifically, VB assumes that a pair member's treatment status (say, \( D_A \)) is weakly higher when both the unit's and the peer's instruments are activate (\( Z_A = 1 \), \( Z_B = 1 \)) compared to only the unit's instrument being activate (\( Z_A = 1 \), \( Z_B = 0 \)). Furthermore, compared to the latter instrument configuration, the value of $D_A$ is weakly smaller when only the peer's instrument is activated (\( Z_A = 0 \), \( Z_B = 1 \)), and weakly smaller still when neither instrument is active (\( Z_A = 0 \), \( Z_B = 0 \)).
In deriving his main results, VB also imposes one-sided non-compliance, which implies the second part of the monotonicity condition above, and facilitates identification of LATE among compliers with nontreated peers as well as the interference effect on untreated units induced by complying peers. 


Importantly, our general results are derived under even weaker monotonicity restrictions on the instruments than in VB. While we also maintain one-sided non-compliance with respect to a treatment's ``own'' instrument, we do not impose any monotonicity restrictions on how the partner instrument affects treatment takeup. As noted in the introduction, we accommodate a non-standard compliance type called cross-defiers, necessary for the application in Example~\ref{ex: grades}, but not present in any other compliance framework we are aware of. In a supplement to this paper, we also investigate relaxations of one-sided noncompliance.

Given our general assumptions, we show that the Wald estimand for treatment $A$, conditional on $Z_B=0$, reflects the average treatment effect in the union of two groups: compliers with $Z_A$ and cross-defiers for $Z_B$. The second Wald estimand, conditional on $Z_B=1$, 
has a complicated interpretation in general. However, under auxiliary conditions this estimand also lends itself to an insightful causal interpretation, given by the weighted average of three local average treatment effects. Similarly to our paper, both BW and VB consider the saturated IV regression 
but under different sets of conditions---BW of course rules out instrument spillovers and assumes statistical independence between the instruments while VB does not allow for cross-defiers and also uses one-sided noncompliance. In BW's framework the coefficient on $D_AD_B$ does identify a meaningful local average interaction effect between the two treatments, while VB does not attempt to draw out any interesting causal parameters buried in this estimand.\footnote{He states a formula in the Supplemental Appendix and only notes that it does not lend itself to a direct causal interpretation.} It is true that in our general framework this coefficient also does not lend itself to a ``clean'' interpretation. Instead, the interaction effect, often of central interest in applications, is bound up with terms that result from instrument spillovers and treatment effect heterogeneity across various compliance types. We state partial identification results that provide bounds for the interaction effect. 


A final paper we must acknowledge here is a recent contribution by Bhuller and Sigstad (2024), which also considers a multiple treatment framework. However, the study differs from ours, as well as from BW and VB, in that it does not focus on identifying a specific LATE or interaction effect within a well-defined group of compliers. Instead, it aims to provide conditions under which a 2SLS regression with multiple treatments consistently estimates the weighted average effect of a given treatment across multiple compliance types, ensuring proper (in particular, non-negative) weights under arbitrary treatment effect heterogeneity. 
The authors demonstrate that it is necessary and sufficient to have an ``average conditional monotonicity'' and ``no cross effects'' condition hold, encompassing the treatment exclusion restriction (along with treatment monotonicity in the own instrument) as a special case. In Section~\ref{sec: identification} below we will use more specific insights from this paper in explaining the structure of our own results. 




\section{A potential outcome framework for pairs}\label{sec: RCM for pairs}

\subsection{Variable definitions}\label{subsec: vardef}

The population consists of ordered pairs of individuals (e.g., married couples); we will refer to the first member of a pair as member $A$ and the second as member $B$. There are two potentially different binary treatments: $D_A$ is targeted at member $A$ and $D_B$ is targeted at member $B$. 
By representing individual units as pairs with identical members, the setup also accommodates the analysis of two (interacting) treatments received by a single unit.


We are interested in the effect of $D_A$ and/or $D_B$ on some dependent variable $Y$. This outcome may be associated with member $A$ alone, member $B$ alone, or the pair itself. The observed value of $Y$ is given by one of four potential outcomes: $Y(d_A,d_B)$ for $d_A,d_B\in\{0,1\}$. For example, $Y(1,0)$ is the potential outcome if one imposes $D_A=1$ and $D_B=0$, i.e., member $A$ is exposed to treatment $A$, but member $B$ is not exposed to treatment $B$. To make the notation less cluttered, we will omit the comma and simply write $Y(10)$ whenever actual figures ('1' and/or '0') are used in the argument. Using the potential outcomes and the treatment status indicators, we can formally express the observed outcome as
\begin{eqnarray}
Y&=&Y(11)D_AD_B+Y(10)D_A(1-D_B)+Y(01)(1-D_A)D_B\nonumber\\
&&+Y(00)(1-D_A)(1-D_B).\label{def: Y}
\end{eqnarray}

Treatment effect identification is facilitated by a pair of binary instruments, $Z_A$ and $Z_B$, assigned to pair members $A$ and $B$, respectively. We think of these instruments as indicators of (randomly assigned) treatment eligibility or the presence of an exogenous incentive to take the corresponding treatment. The leading example is a randomized control trial, where $Z_A$ and $Z_B$ are the experimenter's intended treatment assignments for pair member $A$ and $B$, respectively. Compliance with these assignments is, however, endogenous and possibly coordinated across pair members. We refer to $Z_A$ as member/treatment $A$'s \emph{own instrument} and $Z_B$ as the \emph{partner instrument}. The labels are of coursed reversed for treatment $B$.

Thus, there are four potential treatment status indicators associated with each pair member; they are denoted as $D_A(z_A, z_B)$ for member $A$ and $D_B(z_A, z_B)$ for member $B$, $z_A, z_B\in\{0,1\}$. For example, $D_A(01)$ indicates whether member $A$ of a pair takes up treatment $A$ when they are not assigned ($Z_A=0$) but their partner is assigned to treatment $B$ ($Z_B=1$). The actual treatment status of member $A$ can be written as
\begin{eqnarray}
D_A&=&D_A(11)Z_AZ_B+D_A(10)Z_A(1-Z_B)+D_A(01)(1-Z_A)Z_B\nonumber\\
&&+D_A(00)(1-Z_A)(1-Z_B)\label{def: D_A}.
\end{eqnarray}
There is of course a corresponding formula for $D_B$.

We now formally impose standard IV assumptions on $Z_A$ and $Z_B$.

\begin{assumption}\label{assn: IV}
[IV] (i) Given the values of the treatment status indicators $D_A$ and $D_B$, the potential outcomes do not depend on the instruments $Z_A$ and $Z_B$. (ii) The instruments $(Z_A, Z_B)$ are jointly independent of the potential outcomes and the potential treatment status indicators. (iii) $P(Z_A=1)\in (0,1)$, $P(Z_B=1)\in (0,1)$ and $P(Z_A=Z_B)\in(0,1)$.
\end{assumption}

The exclusion restriction stated in part (i) of Assumption~\ref{assn: IV} is one of the defining properties of an instrument, and it justifies (ex-post) the potential outcomes being indexed by $(d_A,d_B)$ only. Part (ii), known as ``random assignment,'' states that the instrument values  $(Z_A,Z_B)$ are exogenously determined. This assumption holds, by design, in an experimental setting where intended treatment assignments are explicitly randomized. Part (iii) states that the intended treatment assignments follow a $2\times 2$ factorial design, i.e., there is a positive fraction of pairs assigned to each of the following four categories: treatment $A$ alone, treatment $B$ alone, both treatments, or neither treatment. 

We will impose further assumptions on the potential treatment status indicators in Section~\ref{subsec: subpop}.

\subsection{Parameters of interest}

Let $\mcalP$ be a subset of the population of pairs. We define the following treatment effect parameters and notation:
\begin{itemize}

\item $ATE_{A|\bar B}(\mcalP):= E[Y(10)-Y(00)\mid\mcalP]$ denotes the average effect of applying treatment $A$ alone in the subpopulation $\mcalP$. In other words, this is the average effect of treatment $A$ conditional on treatment $B$ being ``turned off'' in the subpopulation $\mcalP$.


\item $ATE_{A|B}(\mcalP):=E[Y(11)-Y(01)\mid\mcalP]$ denotes the average effect of applying both treatments to the subpopulation $\mcalP$ relative to applying treatment $B$ alone; in other words, this is the average effect of treatment $A$ conditional on maintaining treatment $B$.


\item $ATE_{AB}(\mcalP):=E[Y(11)-Y(00)\mid\mcalP]$ denotes the average effect of applying treatment $A$ and $B$ jointly to the subpopulation $\mcalP$ relative to applying no treatment at all.




\end{itemize}

The parameters $ATE_{A|\bar B}(\mcalP)$ and $ATE_{A|B}(\mcalP)$ are called local average conditional effects, or LACEs, by BW, while $ATE_{AB}(\mcalP)$ is called the local average joint effect (LAJE). For a given group $\mathcal{P}$, the difference between the two conditional effects measures the interaction between the two treatments within $\mcalP$, and is hence termed the local average interaction effect (LAIE) by \emph{ibid.} That is, 
\[
LAIE(\mcalP)=ATE_{A|B}(\mcalP)-ATE_{A|\bar B}(\mcalP).
\]
If the LAIE is positive, the two treatments reinforce each other, while if it is negative, then they work against each other. One can define analogous LACE parameters for treatment $B$ by interchanging the roles of $A$ and $B$ in the definitions above. The associated joint and interaction effects stay unchanged.

In case the pairs have distinct members, the interpretation of these parameters also depends on the definition of the outcome $Y$. In particular, if $Y$ is associated with pair member $A$ alone, then $ATE_{A|\bar B}(\mcalP)$ and $ATE_{A|B}(\mcalP)$ measure what is called the direct effect of treatment $A$ by Hudgens and Halloran (2008). On the other hand, if $Y$ is associated with member $B$ alone, then $ATE_{A|\bar B}(\mcalP)$ and $ATE_{A|B}(\mcalP)$ measure the indirect or spillover effect of treatment $A$ on pair member $B$. For example, if the treatment is vaccination, and the outcome is the incidence of a disease, then the vaccination of member $A$ confers protection on member $A$, but also indirectly protects his or her partner.

\subsection{Compliance types}\label{subsec: subpop}

The setup presented in Section~\ref{subsec: vardef} assigns four potential treatment indicators to each pair member, corresponding to the four possible incentive schemes represented by $(Z_A, Z_B)$. Without any further restrictions on treatment takeup, the possible configurations of these 8 potential treatment variables partition the population of pairs into $2^8=256$ different compliance profiles. 
At this level of generality a couple of regression-based estimands can hardly be a meaningful summary of the various average treatment effects across types. Therefore, similarly to VB, we impose one-sided noncompliance with respect to the treatment's own instrument, which dramatically reduces the number of possible compliance profiles. 



\begin{assumption}\label{assn: 1-sided nc, mon}
[One-sided noncompliance] (i) $D_A(0,z)=0$ and (ii) $D_B(z, 0)=0$ for $z\in\{0,1\}$.
\end{assumption}

Assumption~\ref{assn: 1-sided nc, mon} states that neither member of the pair has access to their own treatment unless they have been ``randomized in,'' i.e., the value of their own instrument is 1. In other words, one-sided noncompliance presumes that the experimenter is able to exclude individuals from all sources of the treatment. Whether or not this assumption is reasonable depends on the institutional setting and details of the underlying experiment, but it often fails in practice.  Therefore, we consider relaxations of Assumption~\ref{assn: 1-sided nc, mon} in Appendix \ref{app: relax 1-sided nc}. 

Under Assumption~\ref{assn: 1-sided nc, mon}, each pair member may belong to one of only four compliance types, summarized by the following definition.

\begin{definition} \label{def: compl}
Under Assumption~\ref{assn: 1-sided nc, mon}, member $A$ of a pair $(A,B)$ is one of four compliance types: 
\begin{center}
\begin{tabular}{l c c c c}
     & $D_A(00)$ & $D_A(01)$ & $D_A(10)$ & $D_A(11)$\\
     \hline
     \emph{self-complier (s)} & 0 & 0 & 1 & 1  \\
     \emph{joint complier (j)} & 0 & 0 & 0 & 1 \\
     \emph{never taker (n)}  & 0 & 0 & 0 & 0 \\
     \emph{cross-defier (d)} & 0 & 0 & 1 & 0 \\
     \hline
\end{tabular}\bigskip
\end{center}

%
%

\noindent Furthermore, a pair member is a \emph{complier (c)} if they are either a self-complier or joint-complier. 
\end{definition}

\paragraph{Remarks}

\begin{enumerate}
  \item The corresponding definitions for member $B$ can be obtained by interchanging the two arguments of the potential treatment status indicators, while using the subscript $B$. 

 \item A self-complier's treatment status is determined solely by the value of their own instrument. 
 By contrast, a joint complier takes the treatment if and only if both instruments are turned on; their own instrument is not sufficient to induce participation. (VB's terminology is group compliers.) 

\item Cross-defiers are a non-standard type. If member $A$ is a cross-defier, then they will comply with their own instrument $Z_A$ as long as the other instrument is absent. However, for such individuals the presence of $Z_B$ represents a strong incentive \emph{against} takeup of $D_A$; so strong in fact that it overpowers the presence of $Z_A$ and causes the individual to abandon treatment. Thus, the individual $A$ acts in defiance of $Z_B=1$.\footnote{Defiance is only partial in the the sense that the individual does not necessarily act against $Z_B=0$, but adding this idea to the moniker would be tedious. An alternative label might be ``deserters.''}

\item Finally, a never taker cannot be induced to take the treatment by any instrument configuration.
\end{enumerate}

Given the four individual compliance types, every pair belongs to one of the 16 compliance profiles
$\{s,j,d,n\}\times \{s,j,d,n\}$. For example, $(s,j)$ is the set of pairs where $A$ is a self-complier and $B$ is a joint complier, etc. Furthermore, we will use the notation $(c,\cdot)$ to denote the set of pairs where member $A$ is a complier, etc., and $P(s,n)$ to denote the probability that for a randomly drawn pair $A$ is a self-complier and $B$ is a never-taker, etc.

The following assumption ensures that some of the compliance categories are not vacuous (e.g., there are at least some individuals who respond to their own instrument). 

\begin{assumption}\label{assn: 1st stage}
    [First stage] (i) $P(D_A(10)=1)>0$ and $P(D_B(01)=1)>0$; (ii)  $P(D_A(11)=1)>0$ and $P(D_B(11)=1)>0$.
\end{assumption}

Part (i) of Assumption~\ref{assn: 1st stage} means that $P(s\cup d,\cdot)>0$ and $P(\cdot, s\cup d)>0$ while part (ii) means $P(c,c)>0$. These conditions ensure that the IV estimands considered in Section~\ref{sec: identification} are well defined.



A testable implication of the existence of joint compliers (with respect to treatment $A$) is that if one runs a simple OLS regression of $D_A$ on $Z_B$ in the $Z_A=1$ subsample, then the coefficient of $Z_B$ should be positive. Conversely, if the coefficient is negative, then cross-defiers must be present in the population.\footnote{A zero coefficient means that takeup is consistent with the treatment exclusion restriction, i.e., only the own instrument matters.} 
Clearly, joint compliers treat the presence of $Z_B$ as a positive incentive to take treatment $A$, while cross-defiers treat it as a (strong) disincentive. Given the situation, it may be possible to argue that the two behaviors do not exist simultaneously, i.e., one could rule out joint compliers or cross-defiers for treatment $A$, depending on which type is not needed to explain the sign of the regression coefficient. Our empirical application in Section \ref{sec: application} illustrates how to exploit such simplifications in practice to enhance the interpretation of IV estimands.

Our framework can also accommodate simplifying assumptions on pair formation. For example, one might postulate that there are no $(n,j)$ or $(j,n)$ pairs.
This assumption is plausible if member $A$'s utility of taking treatment $A$ is affected by $Z_B$ only through member $B$'s actual treatment status $D_B$, which more formally means that $D_A(z_A,z_B)$ is of the form $f(z_A, D_B(z_A,z_B))$. If $B$ is a never taker then $D_B=0$, and the value of $Z_B$ is not relevant for $A$'s decision. Hence $A$ cannot be a joint complier. For example, $Z_B$ could be a randomized monetary reward payable only on actual takeup of treatment $B$. If $B$ is never treated, the reward is not paid out, and should be irrelevant to $A$. On the other hand, $(n,j)$ or $(j,n)$ pairs may well exist if $A$ has direct access to the incentive represented by $Z_B$. 
This is the case, for example, if the pair stands for a single unit targeted by two treatments.\footnote{Suppose that individuals participate in a study on the health benefits of physical exercise. Specifically, there are two treatments: running $(D_A)$ and swimming $(D_B)$. The instrument $Z_A$ is a seminar on the health benefits of running and $Z_B$ a seminar on the benefits of swimming. A person who cannot swim will be a never taker with respect to $D_B$. Nevertheless, it is conceivable that for the same person $D_A(10)=0$ but $D_A(11)=1$. This means that a single lecture is not sufficient to convince this person to take up running but after hearing more about the health benefits of exercise, he eventually decides to do so.}


\section{Identification results}\label{sec: identification}

\subsection{Population proportion of compliance profiles}

The exact type of a given pair is generally unobserved as it depends on the pair's behavior in counterfactual scenarios. Nevertheless, the observed conditional probabilities
\begin{equation}\label{compliance probs}
P(D_A=d_A, D_B=d_B\mid Z_A=z_A, Z_B=z_B),\;d_A, d_B, z_A, z_B\in\{0,1\}
\end{equation}
can be used to identify the relative frequency of a number of compliance profiles in the population. Nevertheless, not all probabilities under (\ref{compliance probs}) carry independent information. This is for two reasons: first, for any given $(z_A, z_B)$, the corresponding probabilities add up to 1, and, second, $Z_A=0$ automatically implies $D_A=0$, and $Z_B=0$ automatically implies $D_B=0$ by Assumption~\ref{assn: 1-sided nc, mon}. It follows that there are only five independently informative moments, which of course makes it impossible to identify the relative frequencies of all 16 compliance profiles separately.

Lemma~\ref{lm: prob weights} presents the interpretation of five independent conditional probabilities of the form (\ref{compliance probs}). The subsequent corollary provides further results.
\begin{lemma}\label{lm: prob weights}
Suppose that Assumptions~\ref{assn: IV} and \ref{assn: 1-sided nc, mon} are satisfied. Then:
\begin{eqnarray*}
P(D_A=1\mid Z_A=1, Z_B=0)&=&P(s\cup d, \cdot)=P(s,\cdot)+P(d,\cdot)\\
P(D_B=1\mid Z_A=0, Z_B=1)&=&P(\cdot, s\cup d)=P(\cdot,s)+P(\cdot,d)\\
P(D_A=1, D_B=0\mid Z_A=1, Z_B=1)&=&P(c,n\cup d)=P(c,n)+P(c,d)\\
P(D_A=0, D_B=1\mid Z_A=1, Z_B=1)&=&P(n\cup d,c)=P(n,c)+P(d,c)\\
P(D_A=1, D_B=1\mid Z_A=1, Z_B=1)&=&P(c,c).
\end{eqnarray*}
\end{lemma}

\begin{corollary}\label{cor: prob weights} In consequence, 
\begin{eqnarray*}
P(D_A=0, D_B=0\mid Z_A=1, Z_B=1)&=&P(n\cup d, n\cup d)\\
P(D_A=0\mid Z_A=1, Z_B=0)&=&P(j\cup n,\cdot)=P(j,\cdot)+P(n,\cdot)\\
P(D_A=1\mid Z_A=1, Z_B=1)&=&P(c,\cdot)=P(s,\cdot)+P(j,\cdot)\\
P(D_A=0\mid Z_A=1, Z_B=1)&=&P(n\cup d,\cdot)=P(n,\cdot)+P(d,\cdot)
\end{eqnarray*}
There are three more results that can be obtained by interchanging the roles of $A$ and $B$ in the last three expressions above. 

\end{corollary}

\paragraph{Remarks}

\begin{enumerate}

    \item The proof of Lemma \ref{lm: prob weights} is given in Appendix~\ref{app: proofs main}. 
    

    \item If, say, member $A$ cannot be a cross-defier, then the marginal compliance type probabilities $P(s,\cdot)$, $P(j,\cdot)$ and $P(n,\cdot)$ are all identified. Similarly, if member $A$ cannot be a joint complier, then the marginal compliance probabilities $P(s,\cdot)$, $P(n,\cdot)$ and $P(d,\cdot)$ are identified.

\end{enumerate}

\subsection{The causal interpretation of three IV estimands}\label{subsec: IV estimands}

Each pair in the target population is associated with an observed 5-vector $(Y, D_{A}, D_{B}, Z_{A}, Z_{B})$. Given a sample of observations on this vector, one may run several different IV regressions using the full sample or a suitable subsample.




\begin{enumerate}
    \item [(i)] Consider the IV regression of $Y$ on $D_A$ and a constant in the $Z_{B}=0$ subsample, using $Z_A$ as an instrument for $D_A$. 
    Under general conditions, the Wald estimand
    \begin{equation}\label{Wald: ZB=0}
    \delta_{A0}=\frac{E(Y|Z_A=1, Z_B=0)-E(Y|Z_A=0, Z_B=0)}{E(D_A|Z_A=1, Z_B=0)-E(D_A|Z_A=0, Z_B=0)}.
    \end{equation}
    represents the probability limit of the slope coefficient associated with $D_A$.



    \item [(ii)] The IV regression described in point (i) above can also be implemented in the $Z_{B}=1$ subsample. Under general conditions, the Wald estimand
    \begin{equation}\label{Wald: ZB=1}
    \delta_{A1}=\frac{E(Y|Z_A=1, Z_B=1)-E(Y|Z_A=0, Z_B=1)}{E(D_A|Z_A=1, Z_B=1)-E(D_A|Z_A=0, Z_B=1)}.
    \end{equation}
    represents the probability limit of the slope coefficient associated with $D_A$.

    \item [(iii)] One can also run a full-sample IV regression of $Y$ on a constant, $D_A$, $D_B$, and $D_AD_B$, instrumented by $Z_A$, $Z_B$ and $Z_AZ_B$. More formally, let $D=(D_A, D_B, D_AD_B)'$, $\ddot D=(1, D')'$, $Z=(Z_A, Z_B, Z_AZ_B)'$ and $\ddot Z=(1, Z')'$. Under general conditions, the full-sample IV estimator is a $4\times 1$ vector that converges to the estimand
    \[
    \beta=(\beta_0, \beta_A, \beta_B, \beta_{AB})'=
    [E(\ddot Z\ddot D')]^{-1}E(\ddot Z Y).
    \]


\end{enumerate}

The following three theorems
state the causal interpretation of these estimands. The proofs are provided in Appendix~\ref{app: proofs main}.

\begin{theorem}\label{thm: split Wald ZB=0}
Under Assumptions \ref{assn: IV}, \ref{assn: 1-sided nc, mon} and \ref{assn: 1st stage}, the Wald estimand (\ref{Wald: ZB=0}) satisfies
\[
\delta_{A0}=ATE_{A|\bar B}(s \cup d,\cdot)=ATE_{A|\bar B}(s,\cdot)\frac{P(s,\cdot)}{P(s\cup d,\cdot)}+ATE_{A|\bar B}(d,\cdot)\frac{P(d,\cdot)}{P(s\cup d,\cdot)}.
\]
\end{theorem}

\paragraph{Remarks}

\begin{enumerate}

\item Theorem~\ref{thm: split Wald ZB=0} states that the Wald estimand $\delta_{A0}$ identifies the average effect of treatment $A$ alone among pairs where member $A$ is a self-complier or a cross-defier. If the latter type is not present, the estimand reduces to the ``classic'' LATE parameter. 

\item In some applications $Y$ could be an outcome associated solely with member $B$. In this case $\delta_{A0}$ identifies the average \emph{spillover} effect on member $B$ of a treatment applied to member $A$ --- among pairs where $A$ is a self-complier or cross-defier.

\item Given that $Z_B=0$, $Z_A$ has no effect on $D_B$ because one-sided noncompliance forces $D_B=0$. Furthermore, $Z_A$ has a (weakly) positive effect on $D_A$ since member $A$ cross-defiers also comply with $Z_A$ when $Z_B=0$. These facts in our framework correspond to the ``no cross effects'' and ``average conditional monotonicity'' requirements by Bhuller and Sigstad (2024). As $ibid.$ show, it is precisely under these conditions that IV regression coefficients in a multiple treatment setting recover a properly weighted average treatment effect across compliance types, just as in Theorem~\ref{thm: split Wald ZB=0}.



\end{enumerate}

In Appendix~\ref{app: relax 1-sided nc} we consider an extension of Theorem~\ref{thm: split Wald ZB=0} to the case in which $Z_B$ continues to satisfy one-sided noncompliance but $Z_A$ only obeys a general monotonicity condition. The result is similar to Theorem~\ref{thm: split Wald ZB=0} except that an additional compliance profile must be added to the set of pairs $(s\cup d,\cdot)$; namely, pairs where member $A$ is a ``cross-complier'' in the sense that they take the treatment in response to any one of the instruments being turned on (possibly the partner instrument $Z_B$ alone). The extension is in fact derived from a very general representation theorem that gives causal interpretations to $\delta_{A0}$ and $\delta_{A1}$ without imposing \emph{any} monotonicity conditions on the instruments. This latter result is rather too general to be useful in practice; its main value lies in the fact that it can readily be ``customized'' via auxiliary restrictions that fit the application at hand. For example, the general result also shows that the extension of Theorem~\ref{thm: split Wald ZB=0} to the case in which $Z_B$ does not satisfy one-sided noncompliance is much more complicated, even when $Z_A$ does obey this restriction. 


The next result states the causal interpretation of $\delta_{A1}$. 

\begin{theorem}\label{thm: split Wald ZB=1}
Under Assumption \ref{assn: IV}, \ref{assn: 1-sided nc, mon} and \ref{assn: 1st stage}, the Wald estimand (\ref{Wald: ZB=1}) satisfies
\begin{eqnarray}
\delta_{A1}&=&ATE_{A|\bar B}(c,\cdot)+ATE_{B|\bar A}(\cdot,j)\frac{P(\cdot,j)}{P(c,\cdot)}-ATE_{B|\bar A}(\cdot,d)\frac{P(\cdot,d)}{P(c,\cdot)}\nonumber\\
&+&LAIE(c,c)\frac{P(c,c)}{P(c,\cdot)}\label{Wald ZB=1 interp gen-1}\\
&=&ATE_{AB}(c,j)\frac{P(c,j)}{P(c,\cdot)} + ATE_{A|B}(c,s)\frac{P(c,s)}{P(c,\cdot)}+ATE_{A|\bar B}(c,n\cup d)\frac{P(c,n\cup d)}{P(c,\cdot)}\nonumber\\
&-&ATE_{B|\bar A}(\cdot,d)\frac{P(\cdot,d)}{P(c,\cdot)}+ATE_{B|\bar A}(n\cup d,j)\frac{P(n\cup d,j)}{P(c,\cdot)}\label{Wald ZB=1 interp gen}
\end{eqnarray}
\end{theorem}

\begin{corollary}\label{cor: split Wald ZB=1}
Suppose, in addition, that there are no cross-defiers with respect to either instrument and there are no $(n,j)$ pairs. Then the Wald estimand (\ref{Wald: ZB=1}) is equal to
\begin{equation}\label{Wald ZB=1 interp spec}
\delta_{A1}=ATE_{AB}(c,j)\frac{P(c,j)}{P(c,\cdot)} + ATE_{A|B}(c,s)\frac{P(c,s)}{P(c,\cdot)}+
ATE_{A|\bar B}(c,n)\frac{P(c,n)}{P(c,\cdot)}.
\end{equation}
\end{corollary}

\paragraph{Remarks}

\begin{enumerate}

\item 
In Appendix~\ref{app: relax 1-sided nc} we present an extension of Theorem~\ref{thm: split Wald ZB=1} to the case in which $Z_A$ continues to satisfy  one-sided noncompliance but $Z_B$ only obeys a weaker monotonicity condition.


\item Again, the interpretation of Theorem \ref{thm: split Wald ZB=1} and Corollary~\ref{cor: split Wald ZB=1} is enriched by the fact that $Y$ can be an outcome associated with $A$ alone, $B$ alone, or the pair $(A,B)$.

\item Given that $Z_B=1$, there is a much richer set of possible responses to $Z_A$, making the interpretation of $\delta_{A1}$ complicated. For example,  
$Z_A$ now affects the takeup of $D_B$, positively for member $B$ joint compliers and negatively for cross-defiers. 
Again, the high-level results by Bhuller and Sigstad (2024) show that in this case IV estimands in a multiple treatment setting generally conflate the effects of the various treatments with partly negative weights, just as in Theorem~\ref{thm: split Wald ZB=1}.   

\item The simplifying assumptions imposed in Corollary~\ref{cor: split Wald ZB=1} are motivated and discussed in Section~\ref{subsec: subpop}.

\end{enumerate}


To understand the causal effects appearing in the special case (\ref{Wald ZB=1 interp spec}), consider changing $Z_A$ from 0 to 1 conditional on $Z_B=1$. In this case $(c,j)$ pairs will switch from no treatment at all to both treatments, contributing the first term in (\ref{Wald ZB=1 interp spec}). For $(c,s)$ pairs, member $A$ switches from no treatment to treatment $A$, while member $B$ continues to take treatment $B$ throughout. This contributes the second term. Finally, among $(c,n)$ pairs member $A$ switches from no treatment to treatment $A$, while member $B$ continues to abstain from treatment. This option contributes the last term. As in the absence of cross-defiers $P(c,\cdot)=P(c,j)+P(c,s)+P(c,n)$, the probability weights in (\ref{Wald ZB=1 interp spec}) sum to one, and are identified from the observed data using Lemma~\ref{lm: prob weights} and Corollary~\ref{cor: prob weights}. 
The general expression (\ref{Wald ZB=1 interp gen}) follows the same logic --- it reflects the reaction of various types of pairs to changing $Z_A$ from 0 to 1 while maintaining $Z_B=1$. However, there are now more possibilities, including member $B$ cross-defiers dropping $D_B$ when $Z_A$ is turned on. The end result is a linear combination of average treatment effects where some of the weights are negative and do not sum to one.

Finally, Theorem~\ref{thm: full sample IV} states the causal interpretation of the elements in the coefficient vector $\beta=(\beta_0,\beta_A,\beta_B,\beta_{AB})'$.

\begin{theorem}\label{thm: full sample IV}
Under Assumptions \ref{assn: IV}, \ref{assn: 1-sided nc, mon} and \ref{assn: 1st stage},
\begin{eqnarray}
\beta_0&=&E[Y(00)]\nonumber\\
\beta_A&=&ATE_{A|\bar B}(s\cup d,\cdot)\nonumber\\
\beta_B&=&ATE_{B|\bar A}(\cdot,s\cup d)\nonumber\\
\beta_{AB}&=&ATE_{A|B}(c,c)-ATE_{A|\bar B}(c,c)\label{full IV term1}\\
&+&\frac{P(j,\cdot)}{P(c,c)}[ATE_{A|\bar B}(j,\cdot)-ATE_{A|\bar B}(s\cup d,\cdot) ]\label{full IV term2}\\
&+&\frac{P(\cdot,j)}{P(c,c)}[ATE_{B|\bar A}(\cdot,j)-ATE_{B|\bar A}(\cdot,s\cup d)]\label{full IV term3}\\
&+&\frac{P(d,\cdot)}{P(c,c)}[ATE_{A|\bar B}(s\cup d,\cdot)-ATE_{A|\bar B}(d,\cdot) ]\label{full IV term sd1}\\
&+&\frac{P(\cdot,d)}{P(c,c)}[ATE_{B|\bar A}(\cdot,s\cup d)-ATE_{B|\bar A}(\cdot,d)]\label{full IV term sd2}.
\end{eqnarray}
\end{theorem}

\paragraph{Remarks}

\begin{enumerate}


\item The coefficients on the stand-alone treatment dummies are the same as the split-sample Wald estimands that condition on the partner instrument being zero. 


\item The coefficient on the interaction term has a complex interpretation. Term (\ref{full IV term1}) is the local average interaction effect (LAIE) of the two treatments among $(c,c)$ pairs, which would presumably be of interest in many applications. However, this quantity is confounded by additional terms that depend on the heterogeneity of the average treatment effect across types. For example, term (\ref{full IV term2}) compares the average effect of treatment $A$, applied in isolation, across two subpopulations: pairs where member $A$ is a joint complier versus pairs where member $A$ is a self-complier or a cross-defier. While the latter average treatment effect is identified by $\beta_A$, the former is not. Terms (\ref{full IV term3}), (\ref{full IV term sd1}) and (\ref{full IV term sd2}) have analogous interpretations.

\item The presence of the confounding ``heterogeneity terms'' is due to joint compliers and cross-defiers --- interactive types that react to their partner's instrument as well. Under the treatment exclusion restriction these types are not present, and $\beta_{AB}$ identifies the LAIE of the two treatments among $(s,s)$ pairs, as also shown by Theorem 2 of Blackwell (2017).\footnote{This theorem then holds even without one-sided non-compliance.}

\end{enumerate}

\subsection{Partial identification of LAIE}\label{subsec: partial id}

The coefficient $\beta_{AB}$ in Theorem~\ref{thm: full sample IV} does not have a clean interpretation because it conflates the interaction between the two treatments 
with the heterogeneity of the treatment effects across various compliance types. 
We now show that it is still possible to learn about $LAIE(c,c)$ through bounds constructed under some auxiliary conditions. There are two different approaches. 
First, it is possible to bound $LAIE(c,c)$ directly, based on the moments in its definition. Second, one can take the causal interpretation of $\beta_{AB}$ in Theorem~\ref{thm: full sample IV} as a starting point, and bound the influence of the heterogeneity terms (\ref{full IV term2}) through (\ref{full IV term sd2}). In doing so, one obtains an indirect bound on $LAIE(c,c)$ as well. We present the direct bounds here in the main text; the indirect ones are stated in Appendix \ref{app: partial id indirect}. The application in Section~\ref{sec: application} illustrates both approaches. 

There are four conditional means 
involved in the definition of $LAIE(c,c)$:
\begin{equation}
E[Y(11)|(c,c)],\; E[Y(00)|(c,c)],\; E[Y(10)|(c,c)]\text{ and }E[Y(01)|(c,c)].\label{bdd cond moments}
\end{equation}
The first quantity under (\ref{bdd cond moments}) is identified directly from the data by the conditional expectation of $Y$ given $D_A=D_B=1$ and $Z_A=Z_B=1$; see Lemma \ref{lm: cond mom id} in Appendix A. We bound the remaining moments in the spirit of Manski (1989, 1990), using the following assumption. 

\begin{assumption}\label{assn: bdd outcomes}
[bounds] $(i)$ $Y(d_A,d_B)\in [0,K]$ for some $K>0$; $(ii)$ $Y(10)\ge Y(00)$ and $Y(01)\ge Y(00)$.\medskip
\end{assumption}

Part $(i)$ states that the potential outcomes are bounded; the fact that the lower bound is set to zero is a normalization. Part $(ii)$ postulates that when treatment $A$ is applied in isolation, it has a positive effect on any individual unit, and the same is assumed about treatment $B$. While researchers often hold prior expectations about the sign of an average treatment effect, the requirement that the sign applies uniformly in the population is a non-trivial homogeneity restriction known as monotone treatment response (Manski (1997)). Importantly, however, part $(ii)$ does not restrict the sign of the interaction effect. 

As a first step toward bounding the interaction effect, we provide bounds for the joint effect of the two treatments. 

\begin{theorem}\label{thm: bounds joint}
Suppose that Assumptions \ref{assn: IV} through \ref{assn: bdd outcomes} are satisfied. Then the following inequalities hold true:

\begin{itemize}

    \item [$(a)$] $L_{00}(c,c)\le E[Y(00)|(c,c)]\le U_{00}(c,c)$, where
    \begin{eqnarray*}
        U_{00}(c,c)&=&E[Y(00)]\frac{1}{P(c,c)}-E[Y(00)|(n\cup d,n\cup d)]\frac{P(n\cup d,n\cup d)}{P(c,c)},\\
        L_{00}(c,c)&=& U_{00}(c,c)-
        E[Y(10)|(c,n\cup d)]\frac{P(c,n\cup d)}{P(c,c)}-E[Y(01)|(n\cup d,c)]\frac{P(n\cup d,c)}{P(c,c)},
    \end{eqnarray*}
    and each probability and expectation in the definition of $L_{00}(c,c)$ and $U_{00}(c,c)$ is identified from the data as specified by Lemma~\ref{lm: prob weights} and Lemma~\ref{lm: cond mom id}. 

    \item [$(b)$] In consequence,
    \begin{equation}\label{eqn: ATE_AB bounds}
        E[Y(11)|(c,c)]-U_{00}(c,c)\le ATE_{AB}(c,c) \le E[Y(11)|(c,c)]-L_{00}(c,c),
    \end{equation}
    where $E[Y(11)|(c,c)]$ is identified as in Lemma~\ref{lm: cond mom id}.

\end{itemize}
    
\end{theorem}

\paragraph{Remarks}

\begin{enumerate}

    \item Lemma~\ref{lm: cond mom id} in Appendix~\ref{app: cond mom id} provides the causal interpretation of all conditional moments $E[Y\mid D_A=d_A, D_B=d_B, Z_A=z_A, Z_B=z_B]$, $d_A,d_B,z_A,z_B\in\{0,1\}$.

    \item The proof of Theorem~\ref{thm: bounds joint} is given in Appendix~\ref{app: proofs main}; the construction is similar to Manski's classic work cited above. We expand $E[Y(00)]$ as a weighted average of $E[Y(00)|(c,c)]$ and three other conditional expectations over different subgroups. One of the latter expectations is point-identified from the data, and, under Assumption \ref{assn: bdd outcomes}, the other two expectations can be bounded by identified ones (from above) and zero (from below). We rearrange the resulting inequalities to obtain bounds for $E[Y(00)|(c,c)]$.

    \item If $L_{00}(c,c)$ is negative, it may be replaced by zero; if $U_{00}(c,c)$ is greater than $K$, it may be replaced by $K$. 
    If no such replacements are made, then one can in principle apply the classic theory in  Imbens and Manski (2004) to construct confidence intervals for the partially identified parameter $E[Y(00)|(c,c)]$. In this paper we focus on identification and forego further discussion of inference. 

    \item If one strengthens Assumption \ref{assn: bdd outcomes} to include the condition $Y(11)\ge Y(00)$, then an improved upper bound to $E[Y(00)|(c,c)]$ is given by the minimum of $U_{00}(c,c)$ and $E[Y(11)|(c,c)]$, and the joint treatment effect cannot be less than zero. This extra assumption is not entirely trivial but can still be plausible in applications (see Section~\ref{subsec:appl dir bds}).

\end{enumerate}

The relationship between the average joint and interaction effect can be written as 
\begin{equation}\label{eqn: inter-joint}
LAIE(c,c)=ATE_{AB}(c,c)-ATE_{A|\bar B}(c,c)-ATE_{B|\bar A}(c,c).
\end{equation}
While the average effects of treatments $A$ and $B$, applied in isolation, are not identified in the $(c,c)$ subgroup, they are identified in subgroups $(s\cup d,\cdot)$ and $(\cdot,s\cup d)$, respectively (see Theorem~\ref{thm: full sample IV}). It is natural to take the identified subgroup effects as a reference point and speculate about other groups on the basis of these. In particular, we may write $ATE_{A|\bar B}(c,c)=\lambda_A\cdot ATE_{A|\bar B}(s\cup d,\cdot)$ for some (unknown) multiplier $\lambda_A\ge 0$, and define $\lambda_B$ similarly for treatment $B$. Combining these expressions with (\ref{eqn: ATE_AB bounds}) and (\ref{eqn: inter-joint}) bounds the interaction effect in terms of $\lambda_A$ and $\lambda_B$. One can then consider various hypotheses about these parameters; for example, it may be reasonable to postulate in a given application that $ATE_{A|\bar B}(c,c)$ is at most three times as large as $ATE_{A|\bar B}(s\cup d,\cdot)$, implying $\lambda_A\in [0,3]$. Or, one may plot those $(\lambda_A,\lambda_B)$ pairs for which the upper bound of LAIE is zero, etc. Boundaries of this type are common in the econometrics literature on sensitivity analysis (e.g., Masten and Poirier 2020, Martinez-Iriarte 2021).
We demonstrate the construction and use of such heuristic bounds in the context of our application in Sections \ref{subsec:appl dir bds} and \ref{subsec:appl indir bds}. 


One can bound the interaction effect in a more formal way by also bounding the last two conditional means under (\ref{bdd cond moments}). For these bounds to be potentially tighter than the interval $[0,K]$, we impose further compliance type restrictions. Motivated by the discussion following Assumption \ref{assn: 1-sided nc, mon} in Section \ref{subsec: subpop}, we assume that treatment $A$ does not admit cross defiers while treatment $B$ does not admit joint compliers. (We will argue that such a restrictions are reasonable in our application, at least as a polar case.) This leads to the following result. 

\begin{theorem}\label{thm: bounds laie}
Suppose that Assumptions \ref{assn: IV} through \ref{assn: bdd outcomes} are satisfied. If, in addition, there are no $(d,\cdot)$ pairs and no $(\cdot,j)$ pairs, then the following inequalities hold true:

\begin{itemize}

    \item [$(a)$] $L_{10}(c,c)\le E[Y(10)|(c,c)]\le U_{10}(c,c)$, where
    \begin{eqnarray*}
    L_{10}(c,c)&=& E[Y(10)|(s,\cdot)]\frac{P(s,\cdot)}{P(c,c)}-E[Y(10)|(c,n\cup d)]\frac{P(c,n\cup d)}{P(c,c)}\\
    U_{10}(c,c)&=&L_{10}(c,c)+K\frac{P(j,\cdot)}{P(c,c)},
    \end{eqnarray*}

    \item [$(b)$] $L_{01}(c,c)\le E[Y(01)|(c,c)]\le U_{01}(c,c)$, where
    \begin{eqnarray*}
    U_{01}(c,c)&=& E[Y(01)|(\cdot,s\cup d)]\frac{P(\cdot,s\cup d)}{P(c,c)}-E[Y(01)|(n,c)]\frac{P(n,c)}{P(c,c)},\\
    L_{01}(c,c)&=& U_{01}(c,c)-K\frac{P(\cdot,d)}{P(c,c)},
    \end{eqnarray*}
    and each probability and expectation in the definition of $L_{10}(c,c)$, $U_{10}(c,c)$, $L_{01}(c,c)$ and $U_{01}(c,c)$ is identified from the data as specified by Lemma~\ref{lm: prob weights} and Lemma~\ref{lm: cond mom id}.

    \item [$(c)$] In consequence,
    \begin{eqnarray*}
        &&E[Y(11)|(c,c)]+L_{00}(c,c)-U_{10}(c,c)-U_{01}(c,c)\\
        &&\qquad\le LAIE(c,c) \le E[Y(11)|(c,c)]+U_{00}(c,c)-L_{10}(c,c)-L_{01}(c,c),
    \end{eqnarray*}
    where $E[Y(11)|(c,c)]$ is identified as in Lemma~\ref{lm: cond mom id}.

\end{itemize}
    
\end{theorem}


\section{Empirical Application}\label{sec: application}

\subsection{Data, compliance patterns, and IV estimates}\label{subsec: data}

In this section, we present an empirical illustration of our theory based on data from the Student Achievement and Retention Project first analysed by Angrist, Lang and Oreopoulos (2009). This program, implemented on a campus in Canada in Fall 2005, randomly assigned two treatments, namely academic services (in the form of tutoring) and financial incentives among first year college students whose high school grade point average was lower than the upper quartile. Tutoring included both access to more experienced students trained to provide academic support, as well as sessions aiming at improving study habits. The financial incentives consisted of conditional cash payments, ranging from 1,000 to 5,000 Canadian Dollars, which were paid out if a student reached a specific average grade target in college, as a function of the grade point average previously attained in high school. In our empirical application, $D_A$ is a binary variable indicating the takeup of any form of tutoring, while $D_B$ is an indicator for signing up to receive financial incentives. We are interested in the impact of these treatments on the average grade at the end of the fall semester, which is our outcome variable $Y$. The latter is measured as a credit-weighted average on a 0-100 grading scale for students taking at least one one-semester course.

The random offer of the treatments in the project was partly overlapping in the sense that some students were invited to either one of the treatments, to both, or neither. In this context, the instruments $Z_A$ and $Z_B$ correspond to binary indicators for being invited (and thus, being eligible) for tutoring and financial incentives. Thus, we are in the previously mentioned special case (covered by our framework) where the very same individual is targeted by up to two distinct treatments, rather than having a pair of individuals that might be targeted by the same or separate treatments.

Treatment takeup $D_A$ and $D_B$ may endogenously differ from the random assignment $Z_A$ and $Z_B$, respectively, because unobserved background characteristics such as personality traits likely drive both the treatment decision and academic performance. For instance, among those students offered tutoring and/or financial incentives, less motivated individuals satisfied with lower exam grades might not be willing to take the treatment(s), regardless of having received an offer or not. Due, in part, to such never takers, not all subjects comply with the random assignment, and the groups taking and not taking the treatment(s) generally differ in terms of outcome-relevant characteristics. On the other hand, among those students not offered the respective treatment, nobody managed to not comply with the assignment and take that treatment anyway. For this reason, non-compliance in our data is one-sided, as postulated in Assumption \ref{assn: 1-sided nc, mon}.

Applying traditional IV approaches (ruling out relaxations of the treatment exclusion restriction),
the findings of Angrist, Lang and Oreopoulos (2009) point to positive effects of financial incentives or combined treatments among females, but not among males. For this reason, our empirical illustration here only focuses on female students, leaving all in all 948 observations. However, for 150 females the outcome is missing, implying that these students did not take any exams in the fall semester. As the missing outcomes indicator is not statistically significantly associated with $Z_A$ or $Z_B$ (with p-values exceeding 20\%), we drop those observations from the sample, leaving us with 798 females for which the outcomes (as well as the treatments and the instruments) are observed.

\begin{table}[!b]
\begin{center}
	\caption{Treatment and outcome means in the sample and by instruments}\label{tab:descr}
\begin{tabular}{l|ccccc}
  \hline \hline
          &  & $Z_A=1$ & $Z_A=0$ & $Z_A=1$ & $Z_A=0$ \\[-5pt]
 Variable & Total sample & $Z_B=1$ & $Z_B=1$ & $Z_B=0$ &  $Z_B=0$\\ 
  \hline
$D_A$ (tutor) & 0.08 & 0.49 & 0.00 & 0.28 & 0.00 \\
  $D_B$ (fin.\ incentive) & 0.22 & 0.81 & 0.93 & 0.00 & 0.00 \\
  $Y$ (GPA) & 63.78 & 66.98 & 65.75 & 63.57 & 62.83 \\
  \hline
  Number of obs. & 798 & 67 & 134 & 116 & 481 \\
   \hline
\end{tabular}
\end{center}
\par
{\footnotesize Notes: Data from the Student Achievement and Retention Project; see Angrist et al.\ (2009). Female students only; those with missing GPA are dropped. GPA is measured on a 0-100 scale.}
\end{table}

Table \ref{tab:descr} provides descriptive statistics for our evaluation sample, namely the treatment and outcome means in the total sample and in the subsamples defined by the instrument values $Z_A$ and $Z_B$. We see that the treatment frequencies observed in the data are consistent with one-sided noncompliance, since the probability of treatment $D_A$ as well as treatment $D_B$ is zero whenever $Z_A$ and $Z_B$, respectively, is zero. 
As a further observation, the average grade ($Y$) is highest among female students receiving both instruments and lowest among those receiving neither. The difference in average outcomes between the two groups is statistically significant at the 1\% level, pointing to a non-zero reduced form effect of the joint instruments $Z_A$ and $Z_B$ on $Y$. Moreover, the average outcome is somewhat higher among students exclusively eligible for financial incentives than among those exclusively eligible for tutoring (but this difference is not statistically significant at the 10\% level).

To present the effect of the instruments on the treatments (the ``first stage''), Tables \ref{tab:probA}, \ref{tab:probB} and \ref{tab:probAB} report conditional probabilities of the first, second, and joint treatments, respectively, and relate them to specific compliance types. In analyzing treatment takeup patterns, we impose the restriction that there are no cross-defier types with respect to the tutor treatment arm. We consider a similar restriction---no joint compliers---with respect to the financial incentive treatment, but we are more agnostic about this condition and impose it selectively in our bounding exercise later on.  
As discussed in Section \ref{subsec: subpop} (and demonstrated by Corollary \ref{cor: split Wald ZB=1}), such additional restrictions can greatly simplify the interpretation of IV estimands.

\begin{table}[!b]
\begin{center}
	\caption{Conditional probabilities of treatment $D_A$}\label{tab:probA}
\begin{tabular}{l|c|c}
\hline\hline
Conditional probability &Estimate& Interpretation if no $(d,\cdot)$ \\
\hline
 $P(D_A=1\mid Z_A=1,Z_B=0)$ & 0.28  & $P(s,\cdot)$  \\
 $P(D_A=1\mid Z_A=1,Z_B=1)$ & 0.49  &
 $P(c,\cdot)=P(j,\cdot)+P(s,\cdot)$ \\
 $P(D_A=1\mid Z_A=1,Z_B=1)$ & &\\
 $-P(D_A=1\mid Z_A=1,Z_B=0)$ & 0.21 & $P(j,\cdot)$\\
 $P(D_A=0\mid Z_A=1,Z_B=1)$ & 0.51  & $P(n,\cdot)$\\
 \hline
\end{tabular}
\end{center}
\par
{\footnotesize Notes: $D_A$ is the tutor treatment. }
\end{table}

Table~\ref{tab:probA} shows the takeup statistics for $D_A$. The absence of cross-defiers 
means that when being eligible for it, nobody is discouraged from actually taking up tutoring services by additionally being offered financial incentives. We see from Table \ref{tab:probA} that the nonexistence of cross-defiers is consistent with the data since our estimates suggest that $P(D_A=1\mid Z_A=1,Z_B=1)-P(D_A=1\mid Z_A=1,Z_B=0)>0$ (statistically significant at the 1\% level). Yet, we emphasize that this is only a necessary, but not a sufficient condition for the absence of cross-defiers, which requires that \emph{everyone} who receives tutoring when eligible for tutoring alone would also receive tutoring when additionally being eligible for financial incentives. This appears plausible if one agrees that, if anything, financial incentives for good grades should encourage (rather than discourage) the takeup of tutoring given that the latter is expected to increase academic performance. In the absence of cross-defiers, the estimated shares of self-compliers (taking tutoring if and only if eligible for it), joint compliers (taking tutoring if and only if eligible for both treatments) and never takers amount to 28\%, 21\%, and 51\%, respectively.

\begin{table}[!b]
\begin{center}
	\caption{Conditional probabilities of treatment $D_B$}\label{tab:probB}
\begin{tabular}{l|c|c}
\hline\hline
Conditional probability &Estimate& Interpretation\\
\hline
 $P(D_B=1\mid Z_A=0,Z_B=1)$ & 0.93 & $P(\cdot, s\cup d)=P(\cdot,s)+P(\cdot,d)$  \\
 $P(D_B=1\mid Z_A=1,Z_B=1)$ & 0.81 &
 $P(\cdot,c)=P(\cdot, s)+P(\cdot, j)$ \\
 $P(D_B=1\mid Z_A=0,Z_B=1)$ &  &\\
 $-P(D_B=1\mid Z_A=1,Z_B=1)$ & 0.12 & $P(\cdot,d)-P(\cdot,j)$\\
 $P(D_B=0\mid Z_A=0,Z_B=1)$ & 0.07 & $P(\cdot, n)+P(\cdot,j)$\\
 \hline
\end{tabular}
\end{center}
{\footnotesize Notes: $D_B$ is the financial incentive treatment. }
\end{table}

Similarly, Table~\ref{tab:probB} shows the takeup statistics for $D_B$. Not surprisingly, 93\% sign up for the conditional payment when eligible for it (and nothing else). However, the estimates also suggest that $P(D_B=1\mid Z_A=1,Z_B=1) < P(D_B=1\mid Z_A=0,Z_B=1)$,  the 12pp difference being statistically significant at the 5\% level. For this inequality to hold, cross-defiers must be present in the population, and the prevalence of joint compliers must be limited. Cross-defiers with respect to treatment $D_B$ (i.e., the instrument $Z_A$) behave oddly in that they accept the financial incentive if this is the only treatment they are eligible for, but they refuse it if they are additionally eligible for tutoring. While it is not clear why the availability of tutoring should be a disincentive for taking the conditional payment, these types are clearly present in the data.

Some of the subsequent analysis is simpler and more informative under the restriction that there are no joint compliers with respect to the financial incentive. Table~\ref{tab:probB} also shows that the \emph{combined} share of never-takers and joint compliers is estimated to be only 7\%. Still, ruling out joint compliers altogether is a rather strong assumption, since this implies that access to tutoring cannot positively affect sign-up decisions for the financial incentive given eligibility for the latter. This would be violated if some individuals judged their chances of obtaining good grades and realizing the financial rewards to be highly dependent on tutoring, so much so that they would not even sign up for the financial incentive without access to tutoring. This behavior actually seems more reasonable than that of the cross-defiers. On the other hand, $P(\cdot,d)-P(\cdot,j)=0.12$, so the higher the share of joint compliers, the higher the share of cross-defiers must be to explain the data. Given the odd behavior of the latter type, one could argue that their assumed prevalence should be as low as possible, which happens when there are no joint compliers. 


There are additional moments of the data which we present in Appendix \ref{app: emp results}. Specifically, Table \ref{tab:probAB} shows 
the joint distribution of $D_A$ and $D_B$, conditional on eligibility for both treatments. These probabilities identify the shares of specific joint compliance profiles in the population, in accordance with Lemma~\ref{lm: prob weights} and Corollary~\ref{cor: prob weights}. For example, 49\% of the female students are estimated to have a $(c,c)$ profile, 
meaning that their takeup behavior is either in line with the randomized eligibility for tutoring and financial incentives, respectively, or they take both treatments when, and only when, they are eligible for both. In addition, Table~\ref{tab:condmeanY} shows the mean outcome (GPA) conditional on all configurations of the treatment dummies and their instruments. As shown by Lemma \ref{lm: cond mom id} in Appendix~\ref{app: cond mom id}, these moments identify the mean outcome in subgroups with various compliance profiles.  


Finally, in Table \ref{tab:IVreg} we provide the results of a two stage least squares regression of $Y$ (GPA) on treatments $D_A$ and $D_B$, as well as the interaction $D_AD_B$, when using $Z_A$, $Z_B$ and $Z_AZ_B$ as instruments. Theorem 3 shows that the constant term provides an estimate for the mean potential outcome $E[Y(00)]$ when receiving no treatment, suggesting that the average grade amounts to 62.83 points when female students neither take up tutoring, nor sign up for financial incentives. In the absence of cross-defiers with respect to $D_A$, the estimate of $\beta_A$ corresponds to the average effect of among self-compliers when $D_B$ is switched off. Thus, among those complying with eligibility for tutoring, receiving tutoring alone increases the average grade by 2.58 points. 
However, this impact is far from being statistically significant at any conventional level, as the p-value (based on heteroscedasticity-robust standard errors) is equal to 55\%.

\begin{table}[tbp]
\begin{center}
	\caption{IV regression of $Y$ on $D_A$, $D_B$ and $D_AD_B$}\label{tab:IVreg}
\begin{tabular}{l|ccc}
\hline\hline
 Variable & Coefficient estimate & Standard error & P-value \\
  \hline
Constant & 62.83 & 0.55 & 0.00 \\
  $D_A$ (tutor) & 2.58 & 4.35 & 0.55 \\
  $D_B$ (fin.\ incentive)& 3.15 & 1.24 & 0.01 \\
  $D_AD_B$ & 0.69 & 5.31 & 0.90 \\
\hline
\end{tabular}
\end{center}
\par
{\footnotesize Notes: $Y$ is GPA on a 0-100 grading scale. The instruments are the randomized treatment eligibility dummies and their interaction.}
\end{table}

The estimate of $\beta_B$ suggests that among those who either (i) comply with their eligibility for financial incentives
(self-compliers) \emph{or} (ii) refuse the financial incentive when both treatments are available 
(cross-defiers), signing up for the financial incentive has a positive effect of 3.15 points (without tutoring). This effect is statistically significant at the 1\% level. In contrast, the estimate of the interaction term $\beta_{AB}$ is small and statistically insignificant. Nevertheless, as Theorem~\ref{thm: full sample IV} shows, this term is not straightforward to interpret; the fact that it is close to zero does not, by itself, imply that their is no interference across the two treatments.

We now illustrate how to employ the results in Section \ref{subsec: partial id} to learn about the local average interaction effect in the context of our application. In these calculations we will ignore standard errors and treat all point estimates as if they were probability limits. We maintain Assumption \ref{assn: bdd outcomes} throughout, but impose type restrictions only as indicated.

\subsection{Direct bounds on the joint effect and LAIE}\label{subsec:appl dir bds}

We start by applying Theorem~\ref{thm: bounds joint}. Using the estimates from Tables~\ref{tab:probAB} and \ref{tab:condmeanY}, we obtain
\begin{eqnarray*}
    U_{00}(c,c)&=&\frac{62.83}{0.49}-70.55\times\frac{0.19}{0.49}=100.87\;\text{ and }\;\\
    L_{00}(c,c)&=&U_{00}(c,c)-64.83\times\frac{0.31}{0.49}=59.85.
\end{eqnarray*}
Given that $E[Y(11)|(c,c)]=66.94$, this yields $ATE_{AB}(c,c)\in[-33.93,7.09]$. Clearly, $U_{00}$ could be replaced by 100, but it may also be replaced by 66.94 under the additional assumption that $Y(11)\ge Y(00)$, i.e., that taking the two treatments jointly cannot hurt anybody's GPA. This improves the bounds for the joint effect to the reasonably tight interval $[0,7.09]$ without requiring any type restrictions.

\paragraph{Heuristic analysis} As suggested in Section \ref{subsec: partial id}, we can parameterize the standalone effects of $D_A$ and $D_B$ in the $(c,c)$ subgroup as $ATE_{A|\bar B}(c,c)=\lambda_A ATE_{A|\bar B}(s\cup d,\cdot)=2.58\lambda_A$
and $ATE_{B|\bar A}(c,c)=\lambda_B ATE_{B|\bar A}(\cdot, s\cup d)=3.15\lambda_B$ for some multipliers $\lambda_A, \lambda_B\ge 0$. Equation (\ref{eqn: inter-joint}) and the tightened bound on the joint effect then gives
\begin{equation}\label{appl: direct LAIE bd adhoc}
  -2.58\lambda_A-3.15\lambda_B \le LAIE(c,c) \le 7.09 -2.58\lambda_A-3.15\lambda_B. 
\end{equation}

\begin{figure}[!t]  

    \begin{center}
    \includegraphics[scale=0.75]{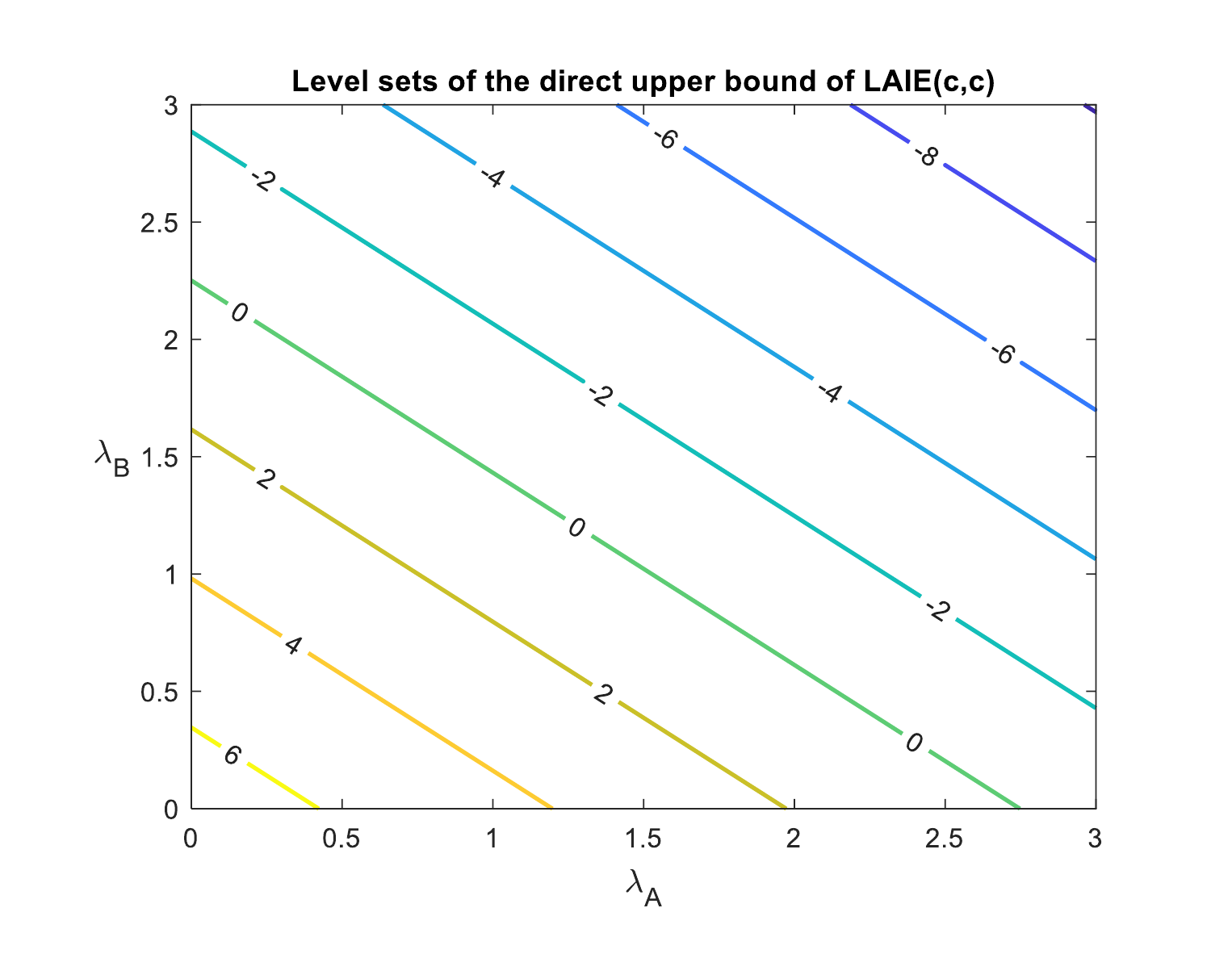}
    \end{center}    
        \par
     \caption{Level sets of the upper bound in (\ref{appl: direct LAIE bd adhoc})}
   
    {\footnotesize Notes: $\lambda_A=ATE_{A|\bar B}(c,c)/ATE_{A|\bar B}(s\cup d,\cdot)$; $\lambda_2=ATE_{B|\bar A}(c,c)/ATE_{B|\bar A}(\cdot, s\cup d)$. }

\label{fig: LAIE ub level zero}
        
\end{figure}

Figure \ref{fig: LAIE ub level zero} depicts the level sets of the upper bound in (\ref{appl: direct LAIE bd adhoc}) for $\lambda_A,\lambda_B \in [0,3]$. Any combination $(\lambda_A,\lambda_B)$ that lies above (i.e., northeast of) the zero line implies a negative LAIE, while for combinations below this line the sign of the interaction effect is unidentified. For example, if the average effect of $D_A$ and $D_B$ is about 25\% percent larger among $(c,c)$ types than among $(s\cup d,\cdot)$ and $(\cdot, s\cup d)$ types, respectively, then LAIE$(c,c)$ is negative. 


\paragraph{Formal analysis} Adopting the restriction that there are no $(d,\cdot)$ and $(\cdot,j)$ pairs and computing the bounds in Theorem \ref{thm: bounds laie} yields $E[Y(10)|(c,c)]\in [37.28,80.14]$ and $E[Y(01)|(c,c)]\in [59.38,83.87]$. Even with the improved upper bound for $E[Y(00)\mid (c,c)]$, the implied LAIE lies in the interval $[-37.21,37.22]$, which is very wide. If we impose the additional assumption that $Y(11)\ge \max\{Y(10),Y(01)\}$, then $E[Y(11)\mid (c,c)]=66.94$ may be used as a tightened upper bound for $E[Y(10)|(c,c)]$ and $E[Y(01)|(c,c)]$ as well.\footnote{The condition $Y(11)\ge \max\{Y(10),Y(01)\}$ implies that, for any individual, the joint effect is (weakly) larger than the standalone effect of each treatment. This assumption is rather strong but it still allows for the interaction effect to be potentially negative; see equation (\ref{eqn: inter-joint}).} This shrinks the bound on the local interaction effect to $[-7.09,37.22]$, but the sign remains unidentified.

\subsection{Indirect bounds on LAIE}\label{subsec:appl indir bds}

We impose the auxiliary condition $P(d,\cdot)=0$ (see Section \ref{subsec: data}),  but for the time being allow for joint compliers in the financial incentive treatment arm ($P(\cdot,j)>0$). The expression for the interaction coefficient $\beta_{AB}$ stated in Theorem~\ref{thm: full sample IV} simplifies, and the local average interaction effect for $(c,c)$ pairs can be expressed as 
\begin{eqnarray}
    &&LAIE(c,c)=ATE_{A|B}(c,c)-ATE_{A|\bar B}(c,c)\nonumber\\
    &&=\beta_{AB}+\frac{P(j,\cdot)}{P(c,c)}ATE_{A|\bar B}(s,\cdot) +\frac{P(\cdot,j)-P(\cdot,d)}{P(c,c)}ATE_{B|\bar A}(\cdot,s\cup d)\label{LAIE id'd terms }\\
    && -\frac{P(j,\cdot)}{P(c,c)}ATE_{A|\bar B}(j,\cdot) +\frac{P(\cdot,d)}{P(c,c)}ATE_{B|\bar A}(\cdot,d)-\frac{P(\cdot,j)}{P(c,c)}ATE_{B|\bar A}(\cdot,j),\label{LAIE unid'd terms }
    \end{eqnarray}
where both average treatment effects under (\ref{LAIE id'd terms }) are identified along with their the probability weights. By contrast, the average treatment effects under (\ref{LAIE unid'd terms }) are not identified and $P(\cdot,j)$ and $P(\cdot,d)$ are also not identified separately. For the identified quantities, we can substitute the point estimates from Tables \ref{tab:probA} through \ref{tab:IVreg} into (\ref{LAIE id'd terms })  and (\ref{LAIE unid'd terms }) to obtain
\begin{eqnarray}
    &&LAIE(c,c)\nonumber\\
    &&=0.69+\frac{0.21}{0.49}(2.58) -\frac{0.12}{0.49}(3.15)\nonumber\\ 
    &&-\frac{0.21}{0.49}ATE_{A|\bar B}(j,\cdot) +\frac{0.12+P(\cdot,j)}{0.49}ATE_{B|\bar A}(\cdot,d) - \frac{P(\cdot,j)}{0.49}ATE_{B|\bar A}(\cdot,j)\nonumber\\
    &&=1.02-0.43 ATE_{A|\bar B}(j,\cdot)+\frac{0.12+P(\cdot,j)}{0.49} ATE_{B|\bar A}(\cdot,d)- \frac{P(\cdot,j)}{0.49}ATE_{B|\bar A}(\cdot,j)\nonumber\\
    \label{LAIE numerical}
\end{eqnarray}



Just as in case of the direct bounds, we can proceed in two ways. 

\paragraph{Heuristic analysis} As suggested in Section \ref{subsec: partial id}, we use the identified local average treatment effects $ATE_{A|\bar B}(s,\cdot)$ and $ATE_{B|\bar A}(\cdot, s\cup d)$ as reference points, and write
\begin{eqnarray*}
  && ATE_{A|\bar B}(j,\cdot)=\lambda_1 ATE_{A|\bar B}(s,\cdot),\; ATE_{B|\bar A}(\cdot,d)=\lambda_2 ATE_{B|\bar A}(\cdot, s\cup d)\quad\text{and}\\
  &&ATE_{B|\bar A}(\cdot,j)=\lambda_3 ATE_{B|\bar A}(\cdot, s\cup d),
\end{eqnarray*}
where the $\lambda_i$, $i=1,2,3$ are scalar multipliers. Using the identified values of $ATE_{A|\bar B}(s,\cdot)$ and $ATE_{B|\bar A}(\cdot, s\cup d)$, and substituting into (\ref{LAIE numerical}) gives
\begin{equation}\label{LAIE: lambda}
    LAIE(c,c)=1.02-1.11\lambda_1+6.43[0.12+P(\cdot,j)]\lambda_2-6.43P(\cdot,j)\lambda_3.
\end{equation}
If it is hypothesized that all the $\lambda_i$ fall into, say, the interval $[0,3]$, then (\ref{LAIE: lambda}) gives the following bounds on the interaction effect:
\[
-2.30-19.29P(\cdot,j)\le ATE_{A|B}(c,c)-ATE_{A|\bar B}(c,c)\le 3.34+19.29P(\cdot,j).
\]
Clearly, the bounds are the tightest when $P(\cdot,j)=0$, i.e., there are no joint compliers with respect to the financial incentive treatment. The sign of the interaction effect is not identified even in this case, which is not too surprising given that $\beta_{AB}$ is so close to zero. 

\begin{figure}[!t]  

    \begin{center}
    \includegraphics[scale=0.75]{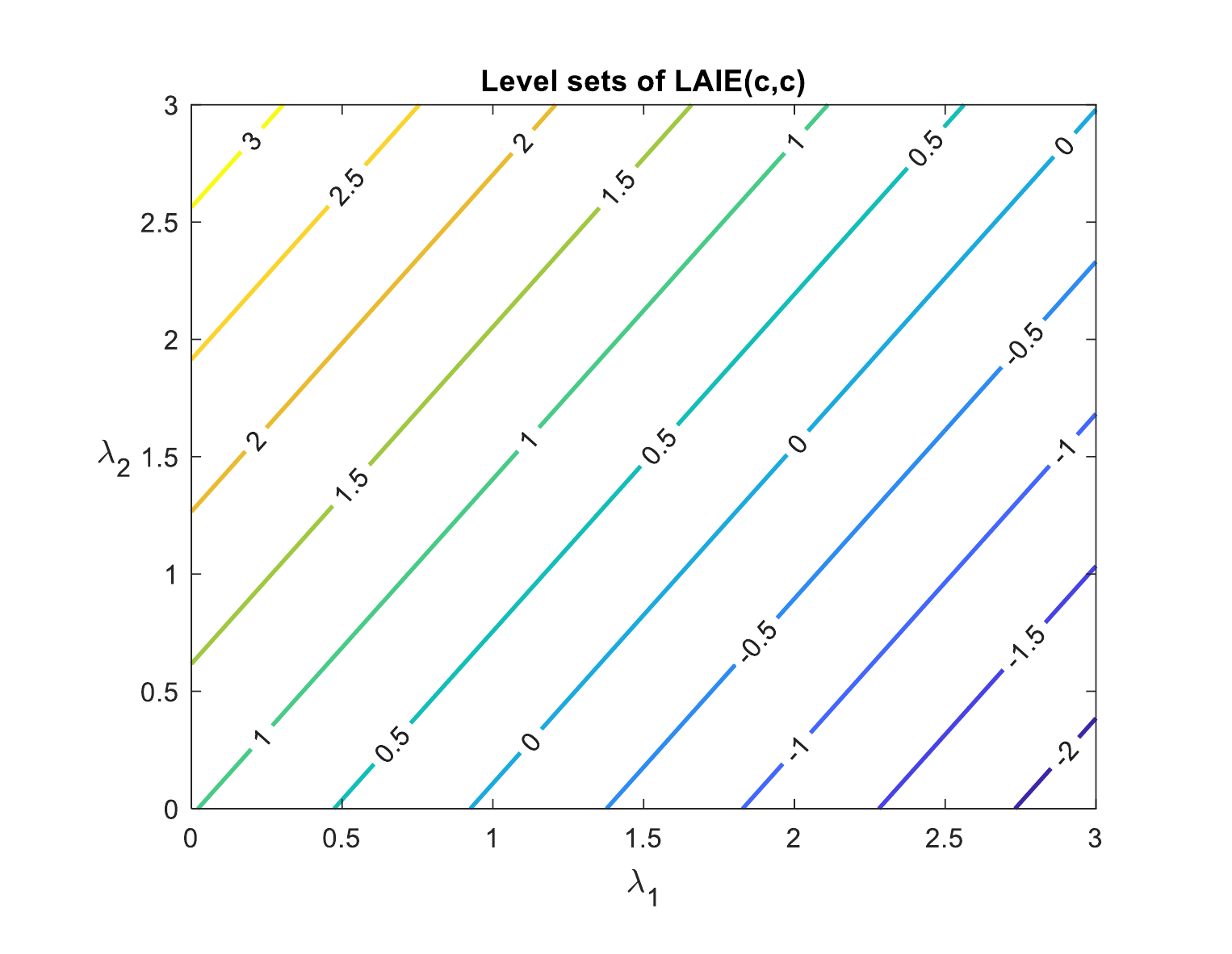}
    \end{center}    
        \par
     \caption{Level sets of $LAIE(c,c)$ as a function of $\lambda_1$ and $\lambda_2$ when $P(\cdot,j)=0$}
   
    {\footnotesize Notes: $\lambda_1=ATE_{A|\bar B}(j,\cdot)/ATE_{A|\bar B}(s,\cdot)$; $\lambda_2=ATE_{B|\bar A}(\cdot,d)/ATE_{B|\bar A}(\cdot, s\cup d)$. No joint compliers with respect to the financial incentive treatment.}

\label{fig: LAIE levels}
        
\end{figure}

If one is willing to work with the assumption that $P(\cdot,j)=0$, expression (\ref{LAIE: lambda}) reduces to a function of $\lambda_1$ and $\lambda_2$ only, and it becomes more straightforward to evaluate $LAIE(c,c)$ in various hypothetical scenarios. In particular, Figure~\ref{fig: LAIE levels} shows the isoquants (level sets) of $LAIE(c,c)$ as a function of $\lambda_1$ and $\lambda_2$. From this graph one can read off the $(\lambda_1,\lambda_2)$ 
pairs that are consistent with, say, a negative interaction between the treatments. 


\paragraph{Formal analysis} Alternatively, one can bound the unknown treatment effects in (\ref{LAIE numerical}) by computing the Manski-type bounds stated in Theorem~\ref{thm: cond moment bounds}, while also imposing $P(\cdot,j)=0$. This yields, after replacing any uninformative bounds with trivial ones, $ATE_{A|\bar B}(j,\cdot)\in [0,43.88]$ and $ATE_{A|\bar B}(\cdot,d)\in [0,100]$. Combining these bounds with equation (\ref{LAIE numerical}) gives LAIE$(c,c)\in[-17.85,25.51]$, which is rather too loose to be useful. One can intersect this interval with the formal direct bounds for LAIE, but the sign remains unidentified. Nevertheless, these Manski-type bounds can still be informative in other applications.

\section{Conclusion}\label{sec: concl}

We study randomized experiments (or quasi-experiments) in which the experimental units are potentially exposed to one of two different treatments, both, or none.  Compliance with the intended treatment assignments, described by two binary instruments, is allowed to be endogenous. 
Our setup allows for the presence of compliance types that, to our knowledge, have not been considered in the literature, but are needed to accommodate some applications. In particular, 
there can be individuals in the population for whom the presence of $Z_B$ (or, resp.\ $Z_A$) represents a \emph{negative} incentive to take treatment $D_A$ (or, resp.\ $D_B$); we call this type cross-defiers. At the same time, we allow for joint compliers as well --- a type that reacts positively to the partner instrument and ultimately takes a given treatment whenever both instruments are present.


We develop the causal interpretation of three IV estimands in our framework. The price of generality is that some of the identification results are weak in the sense that interesting causal parameters are inextricably tied up with terms arising from treatment effect heterogeneity, and auxiliary conditions are needed to obtain more useful interpretations. Alternatively, we provide partial identification results with the goal of bounding the interaction effect between the two treatments, which is frequently of interest in applications. 

A clear advantage of the general approach is that one does not need to pre-commit to a theoretical framework that does not quite fit the data, and any further auxiliary conditions can be tailored to the application at hand. (For example, Blackwell (2017) needs to drop a small set of data points because they violate the treatment exclusion restriction.) Our empirical application, which analyzes a program randomly offering tutoring services (treatment $A$) and financial incentives (treatment $B$) to female college students, illustrates the advantages of starting from a general interpretative framework as well as the use of our partial identification results.

\newpage

\begin{apabib}

	
Angrist, J.D., Lang, D.\ and Oreopoulos, P.\ (2009). ``Incentives and Services for College Achievement: Evidence from a Randomized Trial,'' \emph{American Economic Journal: Applied Economics}, 1, 136-63.


Behaghel, L., Cr\'{e}pon, B.\ and Gurgand, M.\ (2013). ``Robustness of the encouragement design in a two-treatment randomized control trial,'' IZA Discussion Paper No 7447.

Bhuller, M.\ and Sigstad, H.\ (2024). ``2SLS with multiple treatments,'' \emph{Journal of Econometrics},  242, 105785.

Blackwell, M.\ (2017). ``Instrumental Variable Methods for Conditional Effects and Causal Interaction in Voter Mobilization Experiments,'' \emph{Journal of the American Statistical Association}, 112, 590-599.

Ferracci, M., Jolivet, G.\ and van den Berg, G. J. (2014). ``Evidence of treatment spillovers within markets,'' \emph{Review of Economics and Statistics}, 96, 812–823.

Fisher, R.A.\ (1925). \textit{Statistical Methods for Research Workers.} Edinburgh: Oliver and Boyd.



Heckman, J.J.\ and Pinto, R.\ (2018). ``Unordered Monotonicity,''  \emph{Econometrica}, 86, 1-35.

Hong, G.\ and Raudenbush, S.W.\  (2006). ``Evaluating kindergarten retention policy. Journal of
the American Statistical Association,'' \emph{Journal of the American Statistical Association}, 101, 901–910.

Huber, M.\ and Steinmayr, A. (2021). ``A framework for separating individual-level treatment effects from spillover effects,'' \emph{Journal of Business \& Economic Statistics}, 39, 422–436.

Hudgens, M.G.\ and Halloran, M.E.\ (2008). ``Toward Causal Inference with Interference,'' \emph{Journal of the American Statistical Association}, 103, 832-842.

Imai, K.\, Jiang, Z.\  and Malani, A.\ (2021). ``Causal inference with interference and noncompliance in two-stage randomized experiments,'' \emph{Journal of the American Statistical Association}, 116, 632–644.

Imbens, G.W.\ and Angrist, J.D.\ (1994). ``Identification and Estimation of Local Average Treatment Effects,'' \emph{Econometrica}, 62, 467-475.

Imbens, G.W.\ and Manski, C.F.\ (2004). ``Confidence Intervals for Partially Identified Parameters.'' \textit{Econometrica}, 72, 1845-1857. https://doi.org/10.1111/j.1468-0262.2004.00555.x

Goff (2022). ``A Vector Monotonicity Assumption for Multiple Instruments,'' working paper arXiv:2009.00553.

Kang, H.\ and Imbens, G.\ (2016). ``Peer Encouragement Designs in Causal Inference with Partial Interference and Identification of Local Average Network Effects,'' working paper arXiv:1609.04464.

Kirkeboen, L.J., Leuven, E.\ and Mogstad, M. (2016). ``Field of Study, Earnings, and Self-Selection,'' \emph{The Quarterly Journal of Economics}, 131, 1057-1111.

Kormos, M. (2018). ``Paired 2$\times$2 Factorial Design for Treatment Effect Identification and Estimation in the Presence of Paired Interference and Noncompliance,'' MA thesis, Central European University.

Lee, S.\ and Salanié, B. (2018). ``Identifying Effects of Multivalued Treatments.'' \emph{Econometrica}, 86, 1939-1963. https://doi.org/10.3982/ECTA14269

Manski, C.F. (1989). ``The Anatomy of the Selection Problem,'' \emph{Journal of Human Resources}, 24, 343-360.

Manski, C.F. (1990). ``Nonparametric Bounds on Treatment Effects,'' \emph{American Economic Review}, 80, 319-323.

Manski, C.F. (1997). ``Monotone treatment response,'' \emph{Econometrica}, 65, 1311–1334.

Masten, M.A., and A.\ Poirier (2020). ``Inference on Breakdown Frontiers,'' \emph{Quantitative Economics}, 11, 41-111.

Martínez-Iriarte, J.\ (2021). ``Sensitivity analysis in unconditional quantile effects,'' working paper: University of California, Santa Cruz.

Mogstad, M., Torgovitsky, A.\ and Walters, C.R.\ (2020). ``Policy Evaluation with Multiple Instrumental Variables,'' NBER Working Paper 27546".


Rubin, D.B.\ (1974). ``Estimating Causal Effects of Treatments in Randomized and Nonrandomized Studies,'' \emph{Journal of Educational Psychology}, 66, 688-701.

Rubin, D.B.\ (1978). ``Bayesian Inference for Causal Effects,'' \emph{The Annals of Statistics}, 6, 34-58.

Sobel, M.E.\ (2006). ``What Do Randomized Studies of Housing Mobility Demonstrate? Causal Inference in the Face of Interference,'' \emph{Journal of the American Statistical Association}, 101, 1398-1407.


Vazquez-Bare, G.\ (2022). ``Causal Spillover Effects Using Instrumental Variables,'' \emph{Journal of the American Statistical Association}, forthcoming. DOI: 10.1080/01621459.2021.2021920

\end{apabib}

\newpage

\appendix

\numberwithin{lemma}{section}
\numberwithin{theorem}{section}
\numberwithin{table}{section}

\newcounter{asscountapp}
\numberwithin{asscountapp}{section}
\newenvironment{assumptionapp}
{\refstepcounter{asscountapp}\bigskip\noindent\textsc{Assumption~\Alph{section}.\arabic{asscountapp}}}
{\smallskip}

\section*{Appendix}

{\footnotesize

\section{Additional identification results}
\label{app: cond mom id}




\subsection{A simple technical lemma}

\numberwithin{lemma}{section}

The following decomposition follows easily from the law of iterated expectations and is used in deriving several results throughout the paper.

\begin{lemma}\label{lm: ATE decomp}
    Let $\mathcal{P}$ be a set of pairs, i.e., $\mathcal{P}\subseteq \{s,j,n,d\}^2$. Let the nonempty sets $\mathcal{P}_1\subset \mathcal{P} $ and $\mathcal{P}_2\subset \mathcal{P} $ form a partition of $\mathcal{P}$. Then:
    \[
    ATE_{A|B}(\mathcal{P})P(\mathcal{P})=ATE_{A|B}(\mathcal{P}_1)P(\mathcal{P}_1)+ATE_{A|B}(\mathcal{P}_2)P(\mathcal{P}_2).
    \]
The same decomposition holds true if $A|B$ is replaced by $A|\bar B$, $AB$, etc., throughout.
\end{lemma}

\prf We can write
\begin{eqnarray*}
ATE_{A|B}(\mathcal{P})P(\mathcal{P})&=&
E\big[Y(11)-Y(01)\big|\,\mathcal{P}\big]P(\mathcal{P})=E\big\{\big[Y(11)-Y(01)\big]\cdot 1_\mathcal{P}\big\}\\
&=&E\Big\{\big[Y(11)-Y(01)\big]\cdot\big[1_{\mathcal{P}_1}+1_{\mathcal{P}_2}\big]\Big\}\\
&=&E\Big\{\big[Y(11)-Y(01)\big]\cdot 1_{\mathcal{P}_1}\big)\Big\}+E\Big\{\big[Y(11)-Y(01)\big]\cdot 1_{\mathcal{P}_2}\Big\}\\
&=&E\big[Y(11)-Y(01)\big|\,\mathcal{P}_1\big]P(\mathcal{P}_1)+E\big[Y(11)-Y(01)\big|\,\mathcal{P}_2\big]P(\mathcal{P}_2)\\
&=&ATE_{A|B}(\mathcal{P}_1)P(\mathcal{P}_1)+ATE_{A|B}(\mathcal{P}_2)P(\mathcal{P}_2).\text{ \eprf}
\end{eqnarray*}

\subsection{Point identification}

The following lemma states the causal interpretation of the conditional mean of $Y$ given all possible configurations of $(D_A, D_B, Z_A, Z_B)$.

\begin{lemma} \label{lm: cond mom id}
(a) Under Assumptions \ref{assn: IV}, \ref{assn: 1-sided nc, mon} and \ref{assn: 1st stage}, the following conditional moments of the potential outcomes are identified:
\begin{eqnarray*}
E[Y(00)]&=&E[Y\mid D_A=0, D_B=0, Z_A=0, Z_B=0]\\
E[Y(00)\mid (n\cup j, \cdot)]&=&E[Y\mid D_A=0, D_B=0, Z_A=1, Z_B=0]\\
E[Y(00)\mid (n\cup d, n\cup d)]&=&E[Y\mid D_A=0, D_B=0, Z_A=1, Z_B=1]\\
E[Y(10)\mid (s\cup d,\cdot)]&=&E[Y\mid D_A=1, D_B=0, Z_A=1, Z_B=0]\\
E[Y(10)\mid (c,n\cup d)]&=&E[Y\mid D_A=1, D_B=0, Z_A=1, Z_B=1]\\
E[Y(11)\mid (c,c)]&=&E[Y\mid D_A=1, D_B=1, Z_A=1, Z_B=1]
\end{eqnarray*}
There are three additional results that can be obtained by interchanging the roles of pair members $A$ and $B$ in the three non-symmetric expressions above.

(b) In consequence of the results above, the following moment is also identified:
\begin{eqnarray*}
E[Y(00)\mid (s\cup d,\cdot)]&=&\frac{1}{P(s\cup d,\cdot)}\big\{E[Y(00)]-E[Y(00)\mid (n\cup j, \cdot)]P(n\cup j,\cdot)\big\}.
\end{eqnarray*}
An additional result can be obtained by interchanging the roles of pair members $A$ and $B$ above.
\end{lemma}


\paragraph{Proof of Lemma~\ref{lm: cond mom id}} For example, the second equality in Lemma~\ref{lm: cond mom id} part $(a)$ can be obtained as follows. The event $\{D_A=0, D_B=0, Z_A=1, Z_B=0\}$ is equivalent to $\{D_A(10)=0, Z_A=1, Z_B=0\}$. (Formally, substitute $Z_A=1, Z_B=0$ into equation (\ref{def: D_A}) and take one-sided non-compliance into account.) By Definition~\ref{def: compl}, the condition $D_A(10)=0$ means that individual $A$ is either a never taker or a joint complier, while we learn nothing about individual $B$. Therefore,
\begin{eqnarray*}
&&E[Y\mid D_A=0, D_B=0, Z_A=1, Z_B=0]=E[Y(00)\mid D_A(10)=0, Z_A=1, Z_B=0]\\
&&=E[Y(00)\mid D_A(10)=0]=E[Y(00)\mid (n\cup j, \cdot)],
\end{eqnarray*}
where the second equality uses the random assignment assumption. The statement in part $(b)$ follows from the law of iterated expectations; specifically, one can write
\[
E[Y(00)]=E[Y(00)\mid (s\cup d,\cdot)]P(s,\cdot)+E[Y(00)\mid (n\cup j, \cdot)](1-P(s\cup d,\cdot)),
\]
and solve for $E[Y(00)\mid (s\cup d,\cdot)]$. \eprf

\subsection{Partial identification}\label{app: partial id indirect}

Lemma \ref{lm: cond mom id} can be combined with  Assumption~\ref{assn: bdd outcomes} to derive bounds for other conditional means of the form $E[Y(d_A,d_B)\mid\mcalP]$. Here we report bounds for the conditional means in involved in the definition of $ATE_{A|\bar B}(j,\cdot)$ and $ATE_{B|\bar A}(\cdot,d)$ in order to bound these effects in expression (\ref{LAIE numerical}). We do so under the auxiliary assumptions $A\notin d$ and $B\notin j$, which are motivated in Section~\ref{subsec: data}.

\begin{theorem}\label{thm: cond moment bounds}
Suppose that Assumptions \ref{assn: IV}, \ref{assn: 1-sided nc, mon}, \ref{assn: 1st stage} and \ref{assn: bdd outcomes} are satisfied. If, in addition,  there are no $(d,\cdot)$ and $(\cdot,j)$ pairs, then the following inequalities hold true:
\begin{itemize}

    \item [(i)] $L_{00}(j,\cdot)\le E[Y(00)|(j,\cdot)]\le U_{00}(j,\cdot)$, where
    \begin{eqnarray*}
        L_{00}(j,\cdot)&\equiv&E[Y(00)|(n\cup j,\cdot)]\frac{P(n\cup j,\cdot)}{P(j,\cdot)} - E[Y(00)|(n,n\cup d)]\frac{P(n,n\cup d)}{P(j,\cdot)}\\
        &-&E[Y(01)|(n,s)]\frac{P(n,s)}{P(j,\cdot)}\\
        U_{00}(j,\cdot)&\equiv&E[Y(00)|(n\cup j,\cdot)]\frac{P(n\cup j,\cdot)}{P(j,\cdot)} - E[Y(00)|(n,n\cup d)]\frac{P(n,n\cup d)}{P(j,\cdot)}
    \end{eqnarray*}

    \item [(ii)] $L_{10}(j,\cdot)\le E[Y(10)|(j,\cdot)]\le U_{10}(j,\cdot)$, where
    \begin{eqnarray*}
        L_{10}(j,\cdot)&\equiv&E[Y(10)|(c,n\cup d)]\frac{P(c,n\cup d)}{P(j,\cdot)} - E[Y(10)|(s,\cdot)]\frac{P(s,\cdot)}{P(j,\cdot)}\\
        U_{10}(j,\cdot)&\equiv&E[Y(10)|(c,n\cup d)]\frac{P(c,n\cup d)}{P(j,\cdot)} - E[Y(10)|(s,\cdot)]\frac{P(s,\cdot)}{P(j,\cdot)}+K\frac{P(c,c)}{P(j,\cdot)}
    \end{eqnarray*}

    \item [(iii)] $L_{01}(\cdot,d)\le E[Y(01)|(\cdot,d)]\le U_{01}(\cdot,d)$, where
    \begin{eqnarray*}
        L_{01}(\cdot,d)&\equiv&E[Y(01)|(\cdot,s\cup d)]\frac{P(\cdot,s\cup d)}{P(\cdot,d)} - E[Y(01)|(n,s)]\frac{P(n,s)}{P(\cdot,d)}-K\frac{P(c,s)}{P(\cdot,d)}\\
        U_{01}(\cdot,d)&\equiv&E[Y(01)|(\cdot,s\cup d)]\frac{P(\cdot,s\cup d)}{P(\cdot,d)} - E[Y(01)|(n,s)]\frac{P(n,s)}{P(\cdot,d)}
    \end{eqnarray*}

    \item [(iv)] $L_{00}(\cdot,d)\le E[Y(00)|(\cdot,d)]\le U_{00}(\cdot,d)$, where
    \begin{eqnarray*}
        L_{00}(\cdot,d)&\equiv& E[Y(00)|(\cdot, s\cup d)]\frac{P(\cdot, s\cup d)}{P(\cdot,d)}-K\frac{P(\cdot,s)}{P(\cdot,d)}\\
        U_{00}(\cdot,d)&\equiv&E[Y(00)|(\cdot, s\cup d)]\frac{P(\cdot, s\cup d)}{P(\cdot,d)}\\
        &\bigwedge&\Big\{ E[Y(00)|(n, n\cup d)]\frac{P(n, n\cup d)}{P(\cdot,d)}+E[Y(10)|(c,n\cup d)]\frac{P(c, n\cup d)}{P(\cdot,d)}\Big\}
    \end{eqnarray*}

\end{itemize}
Moreover, all bounds stated above can be computed from the data using Lemma~\ref{lm: prob weights}, Corollary~\ref{cor: prob weights} and Lemma~\ref{lm: cond mom id}.

\end{theorem}

\paragraph{Remarks}

\begin{enumerate}
    \item Any negative lower bound can be replaced by 0 and any upper bound greater than $K$ can be replaced by $K$.
    \item The bounds stated in Theorem~\ref{thm: cond moment bounds} can be used to bound $ATE_{A|\bar B}(j,\cdot)$ and $ATE_{B|\bar A}(\cdot,d)$ as follows:
    \begin{eqnarray*}
      L_{10}(j,\cdot)-U_{00}(j,\cdot)  &\le ATE_{A|\bar B}(j,\cdot) \le & U_{10}(j,\cdot)-L_{00}(j,\cdot)\label{ATEA|nB(j,.) formal bd}\\
      L_{01}(\cdot,d)-U_{00}(\cdot,d)  &\le ATE_{B|\bar A}(\cdot,d) \le & U_{01}(\cdot,d)-L_{00}(\cdot,d)\label{ATEB|nA(.,d) formal bd}
    \end{eqnarray*}
    Under Assumption~\ref{assn: bdd outcomes} both effects must lie in the $[0,K]$ interval; the bounds above may or may not improve on this in a given application. 

\end{enumerate}


\paragraph{Proof of Theorem~\ref{thm: cond moment bounds}} The proof is similar to the proofs of Theorems~\ref{thm: bounds joint} and \ref{thm: bounds laie}, which is given in Appendix~\ref{app: proofs main}. To save space, here we only present the calculations that lead to part $(i)$. All other derivations follow a similar logic and are available on request.

Using the law of iterated expectations (or Lemma~\ref{lm: ATE decomp}), we can decompose $E[Y(00)|\,(n\cup j,\cdot)]$ as
\begin{eqnarray*}
&&E[Y(00)|\,(n\cup j,\cdot)]P(n\cup j,\cdot)=
E\big\{ Y(00) [1(A\in n) + 1(A\in j)]\big\}\nonumber\\
&&=E\big\{ Y(00) \big[1(A\in n, B\in n\cup d) + 1(A\in n, B\in c) + 1(A\in j)\big]\big\}\nonumber\\
&&=E[Y(00) 1(A\in n, B\in n\cup d)] + E[Y(00)1(A\in n, B\in c)] + E(Y(00)1(A\in j)]\nonumber\\
&&=E[Y(00)\big|\,(n,n\cup d)]P(n,n\cup d)+E[Y(00)\big|\,(n,c)]P(n,c)+E[Y(00)\big|\,(j,\cdot)]P(j,\cdot)
\end{eqnarray*}
so that 
\begin{eqnarray}
    E[Y(00)|\,(j,\cdot)]&=&
    E[Y(00)|\,(n\cup j,\cdot)]\frac{P(n\cup j,\cdot)}{P(j,\cdot)}-E[Y(00)|\,(n,n\cup d)]\frac{P(n,n\cup d)}{P(j,\cdot)}\label{L00 decomp idd}\\
    &-&E[Y(00)|\,(n,c)]\frac{P(n,c)}{P(j,\cdot)}\nonumber
\end{eqnarray}
as $P(j,\cdot)\neq 0$ by Assumption~\ref{assn: 1st stage}. All conditional expectations on the rhs of (\ref{L00 decomp idd}) can be identified using Lemma~\ref{lm: cond mom id} and the auxiliary assumptions that $A\notin d$ and $B\notin j$. Furthermore, by the same auxiliary assumptions and Lemma/Corollary~\ref{lm: prob weights}, $P(j,\cdot)$ and $P(n\cup j, \cdot)$ can be identified as in Table~\ref{tab:probA}, and $P(n,n\cup d)$ and $P(n,c)=P(n,s)$ can be identified as in Table~\ref{tab:probAB}. Hence, the only unidentified expression on the rhs is the conditional expectation $E[Y(00)|\,(n,c)]$. Nevertheless, Assumption~\ref{assn: bdd outcomes} and $B\notin j$ implies
\begin{equation}\label{bd on EY00nc}
    0\le E[Y(00)|\,(n,c)]=E[Y(00)|\,(n,s)]\le E[Y(01)\mid (n,c)],
\end{equation}
where the upper bound is identified as in Lemma~\ref{lm: cond mom id}, using $A\notin d$. Combining (\ref{bd on EY00nc}) with equation (\ref{L00 decomp idd}) gives
\begin{eqnarray*}
&&E[Y(00)|(n\cup j,\cdot)]\frac{P(n\cup j,\cdot)}{P(j,\cdot)} - E[Y(00)|(n,n\cup d)]\frac{P(n,n\cup d)}{P(j,\cdot)}-E[Y(01)|(n,s)]\frac{P(n,s)}{P(j,\cdot)}\\
&&\le E[Y(00)\mid (j,\cdot)]\\
&&\le E[Y(00)|(n\cup j,\cdot)]\frac{P(n\cup j,\cdot)}{P(j,\cdot)} - E[Y(00)|(n,n\cup d)]\frac{P(n,n\cup d)}{P(j,\cdot)},
\end{eqnarray*}
where the lower bound is the definition of $L_{00}(j,\cdot)$ and the upper bound is $U_{00}(j,\cdot)$. \eprf

}

\section{Relaxing one-sided noncompliance}\label{app: relax 1-sided nc}

{\footnotesize

\subsection{Compliance types under more general conditions}\label{app: gen comp types}

We state a monotonicity condition weaker than one-sided noncompliance.

\begin{assumptionapp}\label{assn: monotonicity} [Monotonicity]
\begin{itemize}
    \item [(i)] $D_A(0,z)\le D_A(1,z)$ for $z=0,1$ and (ii) $D_A(01)\le D_A(10)$
    \item [(iii)] $D_B(z,0)\le D_B(z,1)$ for $z=0,1$ and (iv) $D_B(10)\le D_B(01)$
\end{itemize}
\end{assumptionapp}

Part (i) of Assumption~\ref{assn: monotonicity} replaces one-sided noncompliance with the standard monotonicity condition that $Z_A$, treatment $A$'s own instrument, has a weakly positive effect on taking treatment $A$. Part (iii) is the corresponding condition for instrument/treatment $B$. Parts (ii) and (iv) are less standard as analogous conditions do not arise in single-instrument settings. The inequality in part (ii) means that treatment $A$'s own instrument is more effective in inducing the takeup of treatment $A$ than the partner instrument (this is essentially what justifies calling $Z_A$ treatment $A$'s ``own'' instrument); see also Vazquez-Bare (2022). This requirement is implied by one-sided noncompliance as in this case $D_A(01)=0$. Part (iv) is the corresponding condition for instrument/treatment $B$.

It can be easily verified that under Assumption~\ref{assn: monotonicity}, each pair member may belong to four additional types besides $s$, $n$, $j$, $d$. For member $A$, the definition of these extra types is given in Table~\ref{tbl: new types}. 

\begin{table}[!h]
{\footnotesize
\caption{{\small Additional compliance types under montonicity}}
\begin{center}
\begin{tabular}{l c c c c}
     & $D_A(00)$ & $D_A(01)$ & $D_A(10)$ & $D_A(11)$\\
     \hline
     \emph{always-taker} & 1 & 1 & 1 & 1  \\
     \emph{cross-complier} & 0 & 1 & 1 & 1 \\
     \emph{cross-defier type 2} & 1 & 0 & 1 & 1 \\
     \emph{cross-defier type 3}  & 1 & 0 & 1 & 0 \\
     \hline
\end{tabular}
\label{tbl: new types}
\end{center}
}
\end{table}

The definition of an always-taker is obvious --- this type takes the treatment (here, treatment $A$) for any instrument configuration. Cross-compliers are characterized by the condition that it is enough for the partner instrument to be present to induce treatment takeup---hence the name. The monotonicity conditions in Assumption~\ref{assn: monotonicity}(i) and (ii) then guarantee that cross-compliers participate in the treatment whenever the treatment's own instrument is on. There are two additional types whose behavior is less intuitive. Similarly to standard cross-defiers, for these types the presence of the partner instrument clearly represents a disincentive to take treatment $A$. For ``type 3'' cross-defiers, this negative effect is so strong that it even overpowers the effect of the own instrument, similarly to standard cross-defiers. Whether or not it is reasonable to allow for these additional cross-defier types depends on the application at hand.

\subsection{General causal interpretation of the Wald estimands}\label{app: Wald general decomp}

We now state causal interpretations for the Wald estimands $\delta_{A0}$ and $\delta_{A1}$ (see equations (\ref{Wald: ZB=0}) and (\ref{Wald: ZB=1})) which are completely general in that they do not impose \emph{any} monotinicty conditions on the instruments. We then show how this general result can be specialized under combinations of Assumptions~{\ref{assn: 1-sided nc, mon}} and \ref{assn: monotonicity}.

\begin{theorem}\label{thm: Wald decomp general}
    Suppose that Assumption~\ref{assn: IV} (IV) is satisfied. Fix $z\in\{0,1\}$ and let $\mcalP_i$, $i=1,\ldots, 6$ denote six sets of pairs with compliance profiles satisfying the following conditions:
    \begin{eqnarray*}
        &&\mcalP_1=\{D_A(1,z)=1,D_A(0,z)=0\};\; \mcalP_2=\{D_A(1,z)=0,D_A(0,z)=1\};\\
        &&\mcalP_3=\{D_B(1,z)=1,D_B(0,z)=0\};\; \mcalP_4=\{D_B(1,z)=0,D_B(0,z)=1\};\\
        &&\mcalP_5=\{D_A(1,z)D_B(1,z)=1,D_A(0,z)D_B(0,z)=0\};\; \mcalP_6=\{D_A(1,z)D_B(1,z)=0,D_A(0,z)D_B(0,z)=1\}.
    \end{eqnarray*}
Then
\begin{eqnarray}
&&E(Y\mid Z_A=1, Z_B=z)-E(Y\mid Z_A=0, Z_B=z)\nonumber\\
&&=ATE_{A|\bar B}(\mcalP_1)P(\mcalP_1)-ATE_{A|\bar B}(\mcalP_2)P(\mcalP_2)\nonumber\\
&&+ATE_{B|\bar A}(\mcalP_3)P(\mcalP_3)-ATE_{B|\bar A} (\mcalP_4)P(\mcalP_4)+LAIE(\mcalP_5)P(\mcalP_5)-LAIE(\mcalP_6)P(\mcalP_6).\label{ITT ZB=0 general decomp}
\end{eqnarray}
and   
\begin{equation}
E(D_A\mid Z_A=1, Z_B=z)-E(D_A\mid Z_A=1, Z_B=z)=P(\mcalP_1)-P(\mcalP_2).\label{ITT D ZB=0 general decomp}    
\end{equation}

\end{theorem}

\paragraph{Remarks} 
\begin{enumerate}
   
    
    \item For $z=0$, equation (\ref{ITT ZB=0 general decomp}) is the numerator and equation (\ref{ITT D ZB=0 general decomp}) is the denominator of the Wald estimand (\ref{Wald: ZB=0}). For $z=1$, the same equations correspond to the Wald estimand (\ref{Wald: ZB=1}).

    \item Theorem \ref{thm: Wald decomp general} shows that the Wald estimands associated with treatment $A$ can always be represented as a linear combination of three types of quantities: (i) the average effect of treatment $A$ alone in the subpopulations $\mcalP_1$ and $\mcalP_2$; (ii) the average effect of treatment $B$ alone in the subpopulations $\mcalP_3$ and $\mcalP_4$; (iii) the average interaction effect of the two treatments in the subpopulations $\mcalP_5$ and $\mcalP_6$. However, the types of pairs contained in the sets $\mcalP_i$ will differ according to the monotonicity conditions imposed on the instruments. 

\end{enumerate}

\paragraph{Proof of Theorem~\ref{thm: Wald decomp general}} It is readily verified that the outcome equation (\ref{def: Y}) can be rewritten as
\begin{eqnarray}
    Y&=&Y(00)+[Y(10)-Y(00)]D_A+[Y(01)-Y(00)]D_B\nonumber\\
    &+&\big[Y(11)-Y(01)-(Y(10)-Y(00))\big]D_AD_B
    \label{def: Y2}
\end{eqnarray}
Put $\Delta_{A|\bar B}:=Y(10)-Y(00)$, $\Delta_{B|\bar A}:=Y(01)-Y(00)$ and $\Delta_{A|B}:=Y(11)-Y(01)$. Given $z\in\{0,1\}$, we take the expectation of $Y$ conditional on $Z_A=1$ and $Z_B=z$. For example, the expectation of the second term on the rhs of (\ref{def: Y2}) is 
\begin{eqnarray*}
    E[\Delta_{A|\bar B}D_A|Z_A=1, Z_B=z]&=&E[\Delta_{A|\bar B}D_A(1,z)|Z_A=1, Z_B=z]=E[\Delta_{A|\bar B}D_A(1,z)],
\end{eqnarray*}
where the last equality uses the random assignment assumption (Assumption~\ref{assn: IV}(ii)). Treating the other terms similarly, we obtain
\begin{eqnarray*}
    E[Y|Z_A=1, Z_B=z]&=&E[Y(00)]+E[\Delta_{A|\bar B}D_A(1,z)]+E[\Delta_{B|\bar A}D_B(1,z)]\\
    &+&E[(\Delta_{A|B}-\Delta_{A|\bar B})D_A(1,z)D_B(1,z)].
\end{eqnarray*}
By an analogous argument, 
\begin{eqnarray*}
    E[Y|Z_A=0, Z_B=z]&=&E[Y(00)]+E[\Delta_{A|\bar B}D_A(0,z)]+E[\Delta_{B|\bar A}D_B(0,z)]\\
    &+&E[(\Delta_{A|B}-\Delta_{A|\bar B})D_A(0,z)D_B(0,z)].
\end{eqnarray*}
Subtracting the two conditional expectations yields
\begin{eqnarray}
    &&E[Y|Z_A=1, Z_B=z]-E[Y|Z_A=0, Z_B=z]=E\big\{\Delta_{A|\bar B}[D_A(1,z)-D_A(0,z)]\big\}\nonumber\\
    &&\qquad\qquad+E\big\{\Delta_{B|\bar A}[D_B(1,z)-D_B(0,z)]\big\}+E\big\{(\Delta_{A|B}-\Delta_{A|\bar B})[D_A(1,z)D_B(1,z)-D_A(0,z)D_B(0,z)]\big\}.\nonumber\\
    \label{Wald num decomp}
\end{eqnarray}
Without any further restrictions, $D_A(1,z)-D_A(0,z)$ has three possible values: $-1, 0, 1$. Therefore, we can decompose the first term on the rhs of (\ref{Wald num decomp}) as
\begin{eqnarray*}
E\big\{\Delta_{A|\bar B}[D_A(1,z)-D_A(0,z)]\big\}&=&E\big[\Delta_{A|\bar B}\cdot 1_{\{D_A(1,z)-D_A(0,z)=1\}}\big]-E\big[\Delta_{A|\bar B}\cdot 1_{\{D_A(1,z)-D_A(0,z)=-1\}}\big]\\
&=&E\big[\Delta_{A|\bar B}\big|D_A(1,z)=1,D_A(0,z)=0\big]P[D_A(1,z)=1, D_A(0,z)=0]\\
&-&E\big[\Delta_{A|\bar B}\big|D_A(1,z)=0,D_A(0,z)=1\big]P[D_A(1,z)=0,D_A(0,z)=1]\\
&=&ATE_{A|\bar B}(\mcalP_1)P(\mcalP_1)-ATE_{A|\bar B}(\mcalP_2)P(\mcalP_2),
\end{eqnarray*}
where the last equality uses the definition of $ATE_{A|\bar B}(\cdot)$, $\mcalP_1$ and $\mcalP_2$. Decomposing the remaining two terms in (\ref{Wald num decomp}) in an analogous way yields Theorem~\ref{thm: Wald decomp general}.\eprf

The following result specializes Theorem~\ref{thm: Wald decomp general} and extends Theorem~\ref{thm: split Wald ZB=0} at the same time. 

\begin{theorem}\label{thm: split Wald ZB=0 general}
Let Assumptions~\ref{assn: IV} and \ref{assn: 1st stage} hold. Suppose further that $Z_A$ is a monotone instrument (i.e., Assumptions \ref{assn: monotonicity}(i) and (ii) hold), while $Z_B$ still satisfies one-sided noncompliance (Assumption \ref{assn: 1-sided nc, mon}(ii)). Then the Wald estimand $\delta_{A0}$ is equal to
\[
ATE_{A|\bar B}(\mcalP_1),
\]
where $\mcalP_1$ is the set of all pairs in which member $A$ is either a self-complier, a defier, or a cross-complier. 
\end{theorem}

\paragraph{Proof of Theorem~\ref{thm: split Wald ZB=0 general}} We apply Theorem \ref{thm: Wald decomp general} with $z=0$ and the stated conditions on the instruments. It is clear that if $Z_B$ satisfies one-sided noncompliance, then $\mcalP_3$, $\mcalP_4$, $\mcalP_5$ and $\mcalP_6$ are all empty. Furthermore, $\mcalP_2$ is empty by Assumption \ref{assn: monotonicity}(i). Hence, (\ref{ITT ZB=0 general decomp}) reduces to $ATE_{A|\bar B}(\mcalP_1)P(\mcalP_1)$ and (\ref{ITT D ZB=0 general decomp}) reduces to $P(\mcalP_1)$. Under Assumption~\ref{assn: monotonicity}(i) and (ii), the condition $D_A(10)=1$, $D_A(00)=0$ is consistent with $A$ being a self-complier, cross-defier, or cross-complier.\eprf 

The next result also specializes Theorem~\ref{thm: Wald decomp general} and extends Theorem~\ref{thm: split Wald ZB=1} at the same time. 

\begin{theorem}\label{thm: split Wald ZB=1 general}
Let Assumptions~\ref{assn: IV} and \ref{assn: 1st stage} hold. Suppose further that $Z_A$ satisfies one-sided noncompliance (Assumption~\ref{assn: 1-sided nc, mon}(i)), while $Z_B$ is a monotone instrument (i.e., Assumptions \ref{assn: monotonicity}(iii) and (iv) hold). Then the Wald estimand $\delta_{A1}$ is equal to
\begin{eqnarray*}
ATE_{A|\bar B}(\mcalP_1)+ATE_{B|\bar A}(\mcalP_3)\frac{P(\mcalP_3)}{P(\mcalP_1)}-ATE_{B|\bar A}(\mcalP_4)\frac{P(\mcalP_4)}{P(\mcalP_1)}+LAIE(\mcalP_5)\frac{P(\mcalP_5)}{P(\mcalP_1)},
\end{eqnarray*}
where 
\begin{itemize}
    \item $\mcalP_1$ is the set of all pairs in which member $A$ is a complier ($c$);
    \item $\mcalP_3$ is the set of all pairs in which member $B$ is a joint complier ($j$);
    \item $\mcalP_4$ is the set of all pairs in which member $B$ is a cross-defier ($d$) or is a type 3 cross-defier;
    \item $\mcalP_5$ is the set of all pairs in which member $A$ is a complier $(c)$ and member $B$ is a complier, cross-complier, always taker or type 2 cross-defier.   
\end{itemize}
\end{theorem}


\paragraph{Proof of Theorem~\ref{thm: split Wald ZB=1 general}} We apply Theorem \ref{thm: Wald decomp general} with $z=1$ and the stated conditions on the instruments. It is clear that if $Z_A$ satisfies one-sided noncompliance, then $\mcalP_2$ and $\mcalP_6$ are empty. Hence, (\ref{ITT ZB=0 general decomp}) reduces to the terms involving $\mcalP_1$, $\mcalP_3$, $\mcalP_4$ and $\mcalP_5$, and (\ref{ITT D ZB=0 general decomp}) reduces to $P(\mcalP_1)$. The condition $D(11)=1$, which defines $\mcalP_1$, means that $A$ is a complier (as $Z_A$ obeys one-sided non-compliance). The condition $D_B(11)=1$, $D_B(01)=0$, which defines $\mcalP_3$, means that $B$ is a joint complier. The condition $D_B(11)=0$, $D_B(01)=1$, which defines $\mcalP_4$, means that $B$ is a cross-defier or type 3 cross-defier. Finally, the condition $D_A(11)=1$, $D_B(11)=1$, which defines $\mcalP_5$, means that $A$ is a complier while $B$ is a complier or any one of the ``new'' types except a type 3 cross-defier.\eprf

}

\section{Proofs of results in the main text}\label{app: proofs main}

{\footnotesize

\paragraph{Proof of Lemma \ref{lm: prob weights}} For brevity, here we only show the first claim in Lemma~\ref{lm: prob weights}; all other proofs follow the same scheme. Using the relationship between $D_A$ and the potential treatment status indicators (equation (\ref{def: D_A})) as well as the fact that the instruments are randomly assigned (Assumption~{\ref{assn: IV}), we can write
\begin{eqnarray*}
    &&P[D_A=1\mid Z_A=1,Z_B=0]=P[D_A(10)=1\mid Z_A=1,Z_B=0]=P[D_A(10)=1].
\end{eqnarray*}
By Definition~\ref{def: compl}, the condition $D_A(10)=1$ is consistent with $A$ being a self-complier or a cross-defier and does not restrict the behavior of $B$. Hence, $P[D_A(10)=1]=P(s\cup d,\cdot)$. As self-compliers and cross-defiers are two mutually exclusive groups, $P(s\cup d,\cdot)=P(s,\cdot)+P(d,\cdot)$. \eprf

\paragraph{Proof of Theorem~\ref{thm: split Wald ZB=0}} This is a special case of Theorem~\ref{thm: split Wald ZB=0 general} in Appendix~\ref{app: relax 1-sided nc} when $Z_A$ also satisfies one-sided noncompliance. In this case cross-compliers are eliminated from $\mcalP_1$.\eprf

\paragraph{Proof of Theorem~\ref{thm: split Wald ZB=1}} This is a special case of Theorem~\ref{thm: split Wald ZB=1 general} in Appendix~\ref{app: relax 1-sided nc} when $Z_B$ also satisfies one-sided noncompliance. In this case all ``new'' types presented in Table~\ref{tbl: new types} (see Section~\ref{app: gen comp types}) are eliminated from $\mcalP_1$, $\mcalP_3$, $\mcalP_4$, and $\mcalP_5$. This immediately yields equality (\ref{Wald ZB=1 interp gen-1}) in Theorem~\ref{thm: split Wald ZB=1}. To obtain expression (\ref{Wald ZB=1 interp gen}), observe the following decompositions, which follow from Lemma~\ref{lm: ATE decomp}: 
\begin{eqnarray*}
ATE_{A|\bar B}(c,\cdot)P(c,\cdot)&=&ATE_{A|\bar B}(c,c)P(c,c)+ATE_{A|\bar B}(c,n\cup d)P(c,n\cup d)\\
ATE_{B|\bar A}(\cdot,j)P(\cdot,j)&=&ATE_{B|\bar A}(c,j)P(c,j)+ATE_{B|\bar A}(n\cup d,j)P(n\cup d,j)\\      
ATE_{A|B}(c,c)P(c,c)&=&ATE_{A|B}(c,s)P(c,s)+ATE_{A|B}(c,j)P(c,j).
\end{eqnarray*}
Substituting the definition of $LAIE(c,c)$ into (\ref{Wald ZB=1 interp gen-1}), using the equations above, and observing that $ATE_{AB}(\mcalP)=ATE_{A|B}(\mcalP)+ATE_{B|\bar A}(\mcalP)$ yields (\ref{Wald ZB=1 interp gen}). Corollary~\ref{cor: split Wald ZB=1} follows immediately from imposing the stated simplifying conditions on (\ref{Wald ZB=1 interp gen}).\eprf

\subsection*{Proof of Theorem~\ref{thm: full sample IV}}



\paragraph{Proof outline} The proof of Theorem \ref{thm: full sample IV} is based on relating the two-stage least squares formulation of the IV estimand $\beta$ to the reduced form regression, i.e., the regression of $Y$ on $\ddot Z$. In the reduced form regression the coefficients are intention to treat effects, i.e., they are closely related to the numerators of (\ref{Wald: ZB=0}) and (\ref{Wald: ZB=1}). The causal interpretation of these quantities derive from Theorems \ref{thm: split Wald ZB=0} and \ref{thm: split Wald ZB=1}. On the other hand, the same coefficients can be thought of as the ``product'' of the first stage coefficients (from the regression of $D$ on $\ddot Z$) with the second stage coefficients (from the regression of $Y$ on the predicted values of $D$ and a constant). The second stage coefficients are equal to $\beta$, and hence the causal interpretation of $\beta$ follows from ``dividing'' the reduced form coefficients by the first stage coefficients.

\paragraph{First stage} The first stage of the IV estimator consists of regressing $D_A$, $D_B$ and $D_AD_B$ on $Z=(Z_A, Z_B, Z_A Z_B)'$ and a constant to obtain the linear projections (predicted values) $\hat D_A$, $\hat D_B$ and $\widehat{D_AD_B}$. More specifically, write 
\begin{eqnarray*}
D_A&=&\gamma_{A0} + \gamma_{AA}Z_A + \gamma_{AB}Z_B + \gamma_{A,AB} Z_A Z_B + U_A\\
&=&\gamma_{A0}+\gamma_{A}'Z+U_A\\
D_B&=&\gamma_{B0} + \gamma_{BA}Z_A + \gamma_{BB}Z_B + \gamma_{B,AB} Z_A Z_B + U_B\\
&=&\gamma_{B0}+\gamma_{B}'Z+U_B\\
D_AD_B&=&\gamma_{AB,0} + \gamma_{AB,A}Z_A + \gamma_{AB,B}Z_B + \gamma_{AB,AB} Z_A Z_B + U_{AB}\\
&=&\gamma_{AB,0}+\gamma_{AB}'Z+U_{AB},
\end{eqnarray*}
where $E[U_A]=E[U_B]=E[U_{AB}]=0$, $E[U_AZ]=E[U_BZ]=E[U_{AB}Z]=0$, and 
\[
\gamma_A'=(\gamma_{AA}, \gamma_{AB}, \gamma_{A,AB});\quad\gamma_B'=(\gamma_{BA}, \gamma_{BB}, \gamma_{B,AB});\quad \gamma'_{AB}=(\gamma_{AB,A},\gamma_{AB,B},\gamma_{AB,AB}).   
\]

By standard linear regression theory, the coefficients from the projection of $D_A$ have the following interpretations:
\begin{eqnarray}
\gamma_{A0}&=& E(D_A\mid Z_A=0, Z_B=0)\label{1st stage: gA0}\\
\gamma_{AA}&=& E(D_A\mid Z_A=1, Z_B=0)-E(D_A\mid Z_A=0, Z_B=0) \label{1st stage: gAA}\\
\gamma_{AB}&=& E(D_A\mid Z_A=0, Z_B=1)-E(D_A\mid Z_A=0, Z_B=0) \label{1st stage: gAB}\\
\gamma_{A,AB}&=& E(D_A\mid Z_A=1, Z_B=1)-E(D_A\mid Z_A=0, Z_B=1)\nonumber\\
&-&\big[E(D_A\mid Z_A=1, Z_B=0)-E(D_A\mid Z_A=0, Z_B=0)\big]. \label{1st stage: gAAB}
\end{eqnarray}
Using the IV Assumptions \ref{assn: IV} and \ref{assn: 1-sided nc, mon}, particularly random assignment and one-sided non-compliance w.r.t.\ to the treatment's own instrument, these coefficients reduce to:
\begin{eqnarray}
\gamma_{A0}&=& E[D_A(00)]=0\\
\gamma_{AA}&=& E[D_A(10)]=P(s\cup d,\cdot)\\
\gamma_{AB}&=& E[D_A(01)]-E[D_A(00)]=0\\
\gamma_{A,AB}&=& E[D_A(11)]-E[D_A(10)]=P(c,\cdot)-P(s\cup d,\cdot)=P(j,\cdot)-P(d,\cdot).
\end{eqnarray}
Similar arguments yield
\begin{eqnarray*}
&&\gamma_{B0}=0,\quad \gamma_{BA}=0,\quad \gamma_{BB}=P(\cdot,s\cup d),\quad \gamma_{B,AB}=P(\cdot,j)-P(\cdot,d)\quad\text{ and }\\
&&\gamma_{AB,0}=0,\quad \gamma_{AB,A}=0,\quad \gamma_{AB,B}=0,\quad \gamma_{AB,AB}=P(c,c).
\end{eqnarray*}
Thus, the predicted values from the first stage are simply $\hat D_A=\gamma_A'Z$, $\hat D_B=\gamma_B'Z$ and $\widehat{D_AD_B}=\gamma_{AB}'Z$.

\paragraph{Second stage} The second stage of the IV estimator consists of further projecting $Y$ on $\hat D_A$, $\hat D_B$ and $\widehat{D_AD_B}$. More specifically, write
\begin{eqnarray}
Y&=&\beta_0+\beta_A\hat D_A+\beta_B\hat D_B+\beta_{AB}\widehat{D_AD_B}+U_Y\nonumber\\
&=& \beta_0+\beta_A\gamma_A'Z +\beta_B\gamma_B'Z+\beta_{AB}\gamma_{AB}'Z+U_Y\nonumber\\
&=& \beta_0+(\beta_A, \beta_B, \beta_{AB})\nonumber
\begin{pmatrix}
\gamma_A'Z\\
\gamma_B'Z\\
\gamma_{AB}'Z
\end{pmatrix}
+U_Y\nonumber\\
&=&\beta_0+\beta'\Gamma'Z+U_Y\label{Z on Y 2 stage form},
\end{eqnarray}
where $\beta'=(\beta_A, \beta_B, \beta_{AB})$, $\Gamma'$
is the $3\times 3$ matrix
\[\Gamma'=
\begin{pmatrix}
\gamma_A'\\
\gamma_B'\\
\gamma_{AB}'
\end{pmatrix}
=
\begin{pmatrix}
P(s\cup d,\cdot) & 0 & P(j,\cdot)-P(d,\cdot) \\
0 & P(\cdot, s\cup d) & P(\cdot, j)-P(\cdot,d) \\
0 & 0 & P(c,c)\\
\end{pmatrix},
\]
and 
\begin{equation}\label{red form orth cond}
E(\hat D_AU_Y)=E(\hat D_BU_Y)=E(\widehat{D_AD_B}U_Y)=0. 
\end{equation}

As the diagonal elements of $\Gamma$ are strictly positive under Assumption~\ref{assn: 1st stage}, it follows that $\Gamma'$ is nonsingular or, equivalently, its rows $\gamma_A'$, $\gamma_B'$ and $\gamma_{AB}'$ are linearly independent. Therefore, the orthogonality conditions (\ref{red form orth cond}) can hold only if $E[ZU_Y]=0$. But this means that equation (\ref{Z on Y 2 stage form}) is also the linear projection of $Y$ on $Z$ and a constant, i.e., it can be identified with the reduced form regression
\begin{equation}\label{Y on Z reduced form}
Y=\pi_0+\pi_A Z_A + \pi_B Z_B + \pi_{AB} Z_A Z_B + U_Y.
\end{equation}

\paragraph{Comparison of the reduced form with the second stage} Theorem~\ref{thm: full sample IV}  follows from comparing equation (\ref{Z on Y 2 stage form}) with equation (\ref{Y on Z reduced form}) and using the interpretation of the reduced form regression coefficients as intention to treat effects. In particular, (\ref{Z on Y 2 stage form}) and (\ref{Y on Z reduced form}) imply
\begin{equation}\label{2nd st-red form comp}
\beta'\Gamma'=(\pi_A, \pi_B, \pi_{AB})\Leftrightarrow \beta=\Gamma^{-1}
\begin{pmatrix}
\pi_A\\
\pi_B\\
\pi_{AB}
\end{pmatrix},
\end{equation}
where $\Gamma^{-1}$ is given by
\[
\Gamma^{-1}=
\begin{pmatrix}
\frac{1}{P(s\cup d,\cdot)} & 0 & 0 \\
0 & \frac{1}{P(s\cup d,\cdot)} & 0 \\
-\frac{P(j,\cdot)-P(d,\cdot)}{P(s\cup d,\cdot)P(c,c)} & -\frac{P(\cdot, j)-P(\cdot,d)}{P(\cdot, s\cup d)P(c,c)} & \frac{1}{P(c,c)}
\end{pmatrix}.
\]
The second equation under (\ref{2nd st-red form comp}) then yields
\begin{eqnarray}
\beta_A&=&\frac{\pi_A}{P(s\cup d,\cdot)},\quad \beta_B=\frac{\pi_B}{P(\cdot, s\cup d)},\nonumber\\
\beta_{AB}&=&\frac{1}{P(c,c)}\left[\pi_{AB}-\pi_A\frac{P(j,\cdot)-P(d,\cdot)}{P(s\cup d,\cdot)}-\pi_B\frac{P(\cdot, j)-P(\cdot, d)}{P(\cdot, s\cup d)}\right].\label{beta-pi rel}
\end{eqnarray}

The reduced form coefficients $\pi_A$, $\pi_B$, $\pi_{AB}$ are given by formulas analogous to equations (\ref{1st stage: gAA}), (\ref{1st stage: gAB}) and (\ref{1st stage: gAAB}):
\begin{eqnarray}
\pi_{A}&=& E(Y\mid Z_A=1, Z_B=0)-E(Y\mid Z_A=0, Z_B=0) \label{red form: piAA}\\
\pi_{B}&=& E(Y\mid Z_A=0, Z_B=1)-E(Y\mid Z_A=0, Z_B=0) \label{red form: piB}\\
\pi_{AB}&=& E(Y\mid Z_A=1, Z_B=1)-E(Y\mid Z_A=0, Z_B=1)\nonumber\\
&-&\big[E(Y\mid Z_A=1, Z_B=0)-E(Y\mid Z_A=0, Z_B=0)\big].\label{red form: piAB}
\end{eqnarray}

As can be seen, the coefficient $\pi_A$ is the numerator of the Wald estimand (\ref{Wald: ZB=0}), so that $\pi_A=ATE_{A|\bar B}(s\cup d,\cdot)P(s\cup d, \cdot)$ by Theorem \ref{thm: split Wald ZB=0} and Lemma~\ref{lm: prob weights}. Similarly, $\pi_B=ATE_{B|\bar A}(\cdot, s\cup d)P(\cdot,s\cup d)$. The interpretation of $\pi_{AB}$ is more complicated. The first difference on the rhs of equation (\ref{red form: piAB}) is the numerator of the Wald estimand (\ref{Wald: ZB=1}); hence, 
\[
E(Y\mid Z_A=1, Z_B=1)-E(Y\mid Z_A=0, Z_B=1)=\text{expression }(\ref{Wald ZB=1 interp gen-1})\times P(c,\cdot).
\]
The second difference on the rhs of equation (\ref{red form: piAB}) is the numerator of the Wald estimand (\ref{Wald: ZB=0}); hence, 
\[
E(Y\mid Z_A=1, Z_B=0)-E(Y\mid Z_A=0, Z_B=0)=ATE_{A|\bar B}(s\cup d,\cdot) \times P(s\cup d,\cdot).
\]
Substituting the previous two equations into (\ref{red form: piAB}) yields
\begin{eqnarray}
\pi_{AB}&=&ATE_{A|\bar B}(c,\cdot)+ATE_{B|\bar A}(\cdot,j)P(\cdot,j)-ATE_{B|\bar A}(\cdot,d)P(\cdot,d)\nonumber\\
&+&LAIE(c,c)P(c,c)-ATE_{A|\bar B}(s\cup d,\cdot)P(s\cup d,\cdot)\label{piAB raw}
\end{eqnarray}
Using Lemma~\ref{lm: ATE decomp}, we expand a number of terms in (\ref{piAB raw}) as follows:

\begin{itemize}
    \item $ATE_{A|\bar B}(c,\cdot)P(c,\cdot)=ATE_{A|\bar B}(s,\cdot)P(s,\cdot)+ATE_{A|\bar B}(j,\cdot)P(j,\cdot)$
    
    
    \item $ATE_{A|\bar B}(s\cup d,\cdot)P(s\cup d,\cdot)=ATE_{A|\bar B}(s,\cdot)P(s,\cdot)+ATE_{A|\bar B}(d,\cdot)P(d,\cdot)$



    
\end{itemize}
Substituting the expressions under the bullets above into (\ref{piAB raw}) and rearranging gives
\begin{eqnarray}
    \pi_{AB}&=&[ATE_{A|B}(c,c)-ATE_{A|\bar B}(c,c)]P(c,c)\nonumber\\
    &+&ATE_{A|\bar B}(j,\cdot)P(j,\cdot)-ATE_{A|\bar B}(d,\cdot)P(d,\cdot)\nonumber\\
    &+&ATE_{B|\bar A}(\cdot,j)P(\cdot,j)-ATE_{B|\bar A}(\cdot, d)P(\cdot,d)\label{piAB final}.
\end{eqnarray}
The proof is completed by substituting (\ref{piAB final}) and the causal interpretations of $\pi_A$ and $\pi_B$ into the equations under (\ref{beta-pi rel}). After some straightforward manipulations, this yields the expressions for $\beta_A$, $\beta_B$ and $\beta_{AB}$ stated in Theorem~\ref{thm: full sample IV}. \eprf

\subsection*{Proof of Theorem~\ref{thm: bounds joint}}

$(a)$ Under Assumption~\ref{assn: 1-sided nc, mon}, the set $\{(c,c), (c,n\cup d), (n\cup d, c), (n\cup d, n\cup d)\}$ is a partition of the space of possible pairs. Thus, 
using the law of iterated expectations (or Lemma~\ref{lm: ATE decomp}), we can expand $E[Y(00)]$ as 
\begin{eqnarray}
    E[Y(00)]&=&E[Y(00)\mid (c,c)]P(c,c)+E[Y(00)\mid (n\cup d,n\cup d)]P(n\cup d,n\cup d)\nonumber\\
    &+&E[Y(00)\mid (c,n\cup d)]P(c,n\cup d)+E[Y(00)\mid (n\cup d,c)]P(n\cup d,c)\label{EY00 expand}.
\end{eqnarray}
By Lemma~\ref{lm: cond mom id}, the second conditional expectation on the rhs of equation (\ref{EY00 expand}) is identified as $E[Y|D_A=0, D_B=0, Z_A=1, Z_B=1]$. Under Assumption $\ref{assn: bdd outcomes}$, 
\[
0\le E[Y(00)\mid (c,n\cup d)] \le E[Y(10)|(c,n\cup d)],
\]
where Lemma~\ref{lm: cond mom id} identifies the upper bound as $E[Y|D_A=1,D_B=0, Z_A=1,Z_B=0]$. Similarly, 
\[
0\le E[Y(00)\mid (n\cup d,c)] \le E[Y(01)|(n\cup d,c)],
\]
where Lemma~\ref{lm: cond mom id} identifies the upper bound as $E[Y|D_A=0,D_B=1, Z_A=1,Z_B=0]$. Rearranging equation (\ref{EY00 expand}) to express $E[Y(00)\mid (c,c)]P(c,c)$ and taking the above inequalities into account yields
\begin{eqnarray*}
    &&E[Y(00)]-E[Y(00)\mid (n\cup d,n\cup d)]P(n\cup d,n\cup d)\\
    &&-E[Y(10)\mid (c,n\cup d)]P(c,n\cup d)-E[Y(01)\mid (n\cup d,c)]P(n\cup d,c)\\
    &&\le E[Y(00)\mid (c,c)]P(c,c)\\
    &&\le E[Y(00)]-E[Y(00)\mid (n\cup d,n\cup d)]P(n\cup d,n\cup d).
\end{eqnarray*}
By Assumption~\ref{assn: 1st stage}, $P(c,c)>0$, and dividing through by this quantity leads to the definition of $L_{00}(c,c)$ and $U_{00}(c,c)$. All probabilities appearing in the definition can be identified from Lemma~\ref{lm: prob weights} and Corollary~\ref{cor: prob weights}.

$(b)$ This is immediate from part $(a)$ and the fact that $ATE_{AB}(c,c)=E[Y(11)\mid (c,c)]-E[Y(00)\mid (c,c)]$. By Lemma~\ref{lm: cond mom id}, $E[Y(11)\mid (c,c)]=E[Y|D_A=1, D_B=1, Z_A=1, Z_B=1]$. \eprf

\subsection*{Proof of Theorem~\ref{thm: bounds laie}}

$(a)$ We start by expanding $E[Y(10)|(c,\cdot)]$ in two different ways. First, $\{(c,\cdot)\}=\{(c,c), (c,n\cup d)\}$, so by Lemma~\ref{lm: ATE decomp}, 
\begin{eqnarray*}
    &&E[Y(10)|(c,\dot)]P(c,\cdot)=E[Y(10)|(c,c)]P(c,c)+E[Y(10)|(c,n\cup d)]P(c,n\cup d).
\end{eqnarray*}
Second, $\{(c,\cdot)\}=\{(s,\cdot), (j,\cdot)\}$, so by Lemma~\ref{lm: ATE decomp},
\begin{eqnarray*}
    &&E[Y(10)|(c,\dot)]P(c,\cdot)=E[Y(10)|(s,\cdot)]P(s,\cdot)+E[Y(10)|(j,\cdot)]P(j,\cdot)
\end{eqnarray*}
Equating the two expansions and solving for $E[Y(10)|(c,c)]$ gives
\begin{eqnarray}
    &&E[Y(10)|(c,c)]\nonumber\\
    &&=E[Y(10)\mid (s,\cdot)]\frac{P(s,\cdot)}{P(c,c)}+E[Y(10)\mid (j,\cdot)]\frac{P(j,\cdot)}{P(c,c)}-E[Y(10)|(c,n\cup d)]\frac{P(c,n\cup d)}{P(c,c)},\label{EY10cc expand}
\end{eqnarray}
where $P(c,c)>0$ by Assumption~\ref{assn: 1st stage}. The first conditional expectation on the rhs of (\ref{EY10cc expand}) is identified by Lemma~\ref{lm: cond mom id} as $E[Y|D_A=1, D_B=0, Z_A=1, Z_B=0]$ under the auxiliary assumption that there are no $(d,\cdot)$ pairs. The second conditional expectation on the rhs of (\ref{EY10cc expand}) is bounded between 0 and $K$ given Assumption~\ref{assn: bdd outcomes}. The third conditional expectation on the rhs of (\ref{EY10cc expand}) is again identified by Lemma~\ref{lm: cond mom id} as $E[Y|D_A=1, D_B=0, Z_A=1, Z_B=1]$. All probabilities in equation (\ref{EY10cc expand}) are identified by Lemma~\ref{lm: prob weights} and Corollary~\ref{cor: prob weights} under the auxiliary assumption that there are no $(d,\cdot)$ pairs. The definitions of $L_{10}(c,c)$ and $U_{10}(c,c)$ follow immediately from (\ref{EY10cc expand}) and the bounds imposed on the second term on the rhs. 

$(b)$ We start by expanding  $E[Y(01)|(\cdot,c)]$ and $E[Y(10)|(\cdot,s\cup d)]$. First, $\{(\cdot,c)\}=\{(c,c), (n\cup d,c)\}$, so by Lemma~\ref{lm: ATE decomp},
\begin{eqnarray*}
    &&E[Y(01)|(\cdot,c)]P(\cdot,c)=E[Y(01)|(c,c)]P(c,c)+E[Y(01)|(n\cup d,c)]P(n\cup d,c)
\end{eqnarray*}
Second, also by Lemma~\ref{lm: ATE decomp},
\begin{eqnarray*}
    &&E[Y(01)|(\cdot,s\cup d)]P(\cdot, s\cup d)=E[Y(01)|(\cdot,s)]P(\cdot,s)+E[Y(01)|(\cdot,d)]P(\cdot,d)
\end{eqnarray*}
Under the auxiliary assumption that there are no $(\cdot,j)$ pairs, the only compliers for treatment $B$ are self-compliers, so that $(\cdot,c)=(\cdot,s)$. We can therefore substitute the first equality above into the second and solve for 
$E[Y(01)|(c,c)]=E[Y(01)|(c,s)]$. This gives
\begin{eqnarray}
    &&E[Y(01)|(c,c)]\nonumber\\
    &&=E[Y(01)\mid (\cdot, s\cup d)]\frac{P(\cdot, s\cup d)}{P(c,c)}
    -E[Y(01)\mid (\cdot,d)]\frac{P(\cdot,d)}{P(c,c)}-E[Y(01)|(n\cup d, c)]\frac{P(n\cup d,c)}{P(c,c)},
    \label{EY01cc expand}
\end{eqnarray}
where $P(c,c)>0$ by Assumption~\ref{assn: 1st stage}. The first conditional expectation on the rhs of (\ref{EY01cc expand}) is identified by Lemma~\ref{lm: cond mom id} as $E[Y|D_A=0, D_B=1, Z_A=0, Z_B=1]$. The second conditional expectation on the rhs of (\ref{EY01cc expand}) is bounded between 0 and $K$ given Assumption~\ref{assn: bdd outcomes}. The third conditional expectation on the rhs of (\ref{EY01cc expand}) is again identified by Lemma~\ref{lm: cond mom id} as $E[Y|D_A=0, D_B=1, Z_A=1, Z_B=1]$. All probabilities in equation (\ref{EY10cc expand}) are identified by Lemma~\ref{lm: prob weights} and Corollary~\ref{cor: prob weights} under the auxiliary assumption that there are no $(\cdot,j)$ pairs. The definitions of $L_{01}(c,c)$ and $U_{01}(c,c)$ follow immediately from (\ref{EY01cc expand}) and the bounds imposed on the second term on the rhs. 

$(c)$ This is an immediate consequence of parts $(a)$ and $(b)$ above, Theorem~\ref{thm: bounds joint}$(a)$, and the fact that 
\[
LAIE(c,c)=E[Y(11)|(c,c)]+E[Y(00)|(c,c)]-E[Y(10)|(c,c)]-E[Y(01)|(c,c)].\;\eprf
\]

} 

\section{Additional empirical results}\label{app: emp results}

{\footnotesize

We provide some additional conditional moments of the data set described in Section \ref{subsec: data}. These moments are used in calculating the bounds presented in Sections~\ref{subsec:appl dir bds} and \ref{subsec:appl indir bds}.

\begin{table}[!h]
\begin{center}
	\caption{Conditional joint probabilities of treatments $D_A$ and $D_B$}\label{tab:probAB}\smallskip
\begin{tabular}{l|c|c}
\hline\hline
Conditional probability &Estimate& Interpretation if no  $(d,\cdot)$\\
\hline
 $P(D_A=1,D_B=1\mid Z_A=1,Z_B=1)$ & 0.49 & $P(c,c)$  \\
 $P(D_A=0,D_B=1\mid Z_A=1,Z_B=1)$ & 0.31 & $P(n,c)$ \\
 $P(D_A=1, D_B=0\mid Z_A=1,Z_B=1)$ & 0 & $P(c,n\cup d)$\\
 $P(D_A=0,D_B=0\mid Z_A=1,Z_B=1)$ & 0.19 & $P(n, n\cup d)$\\
 \hline
\end{tabular}
\end{center}
\par
{\footnotesize Notes: $D_A$ is the tutor treatment; $D_B$ is the financial incentive. The interpretations follow from Lemma~\ref{lm: prob weights} and Corollary~\ref{cor: prob weights}.}
\end{table}

\begin{table}[!h]
\begin{center}
	\caption{Conditional means of $Y$}\label{tab:condmeanY}\smallskip
\begin{tabular}{l|c|c}
\hline\hline
Conditional probability &Estimate& Interpretation\\
\hline

$E[Y\mid D_A=0, D_B=0, Z_A=0, Z_B=0]$ & 62.83  & $E[Y(00)]$\\
$E[Y\mid D_A=0, D_B=0, Z_A=1, Z_B=0]$ & 62.90  & $E[Y(00)\mid (n\cup j, \cdot)]$\\
$E[Y\mid D_A=0, D_B=0, Z_A=0, Z_B=1]$ & 65.07  & $E[Y(00)\mid (\cdot, n\cup j)]$ \\
$E[Y\mid D_A=0, D_B=0, Z_A=1, Z_B=1]$ & 70.55  & $E[Y(00)\mid (n\cup d, n\cup d)]$\\
$E[Y\mid D_A=1, D_B=0, Z_A=1, Z_B=0]$ & 65.24  & $E[Y(10)\mid (s\cup d,\cdot)]$\\
$E[Y\mid D_A=0, D_B=1, Z_A=0, Z_B=1]$ & 65.80  & $E[Y(01)\mid (\cdot, s\cup d)]$\\
$E[Y\mid D_A=1, D_B=0, Z_A=1, Z_B=1]$ & n/a    & $E[Y(10)\mid (c,n\cup d)]$\\
$E[Y\mid D_A=0, D_B=1, Z_A=1, Z_B=1]$ & 64.83  & $E[Y(01)\mid (n\cup d,c)]$\\
$E[Y\mid D_A=1, D_B=1, Z_A=1, Z_B=1]$ & 66.94  & $E[Y(11)\mid (c,c)]$\\
\hline
\end{tabular}
\end{center}
\par
{\footnotesize Notes: $Y$ is GPA measured on a 0-100 scale; $D_A$ is the tutor treatment; $D_B$ is the financial incentive. The interpretations follow from Lemma~\ref{lm: cond mom id} in Appendix~\ref{app: cond mom id}.}
\end{table}


}

\end{document}